\documentclass[english,aps,pre,showpacs,floats,twocolumn]{revtex4-1}
\usepackage[T1]{fontenc}
\usepackage[latin9]{inputenc}
\usepackage{textcomp}
\usepackage{amsmath}
\usepackage{amssymb}
\usepackage{graphicx}
\usepackage{esint}

\makeatletter

\newcommand{\lyxdot}{.}

\makeatother

\usepackage{babel}
\begin{document}

\title{Uniform and non-uniform thermal switching of magnetic particles}

\author{D. A. Garanin}

\affiliation{Physics Department, Lehman College and Graduate School, The City
University of New York, 250 Bedford Park Boulevard West, Bronx, NY
10468-1589, U.S.A.}

\date{\today}
\begin{abstract}
The pulse-noise approach to systems of classical spins weakly interacting
with the bath has been applied to study thermally-activated escape
of magnetic nanoparticles over the uniform and nonuniform energy barriers
at intermediate and low damping. The validity of approximating a single-domain
particle by a single spin is investigated. Barriers for a non-uniform
escape of elongated particles for the uniaxial model with transverse
and longitudinal field have been worked out. Pulse-noise computations
have been done for finite magnetic chains. The linear stability of
the uniform barrier state has been investigated. The crossover between
uniform and nonuniform barrier states has been studied with the help
of the variational approach.
\end{abstract}

\maketitle

\section{Introduction\label{sec:Introduction}}

Miniaturization of magnetic elements in spintronics and memory applications
increases the importance of their thermal stabilty and puts the question
of lifetimes of different magnetic states. These lifetimes are usually
related to overcoming energy barriers under the influence of thermal
agitation, that becomes increasingly easy for small sizes of magnetic
particles, eventually leading to superparamagnetism. On the other
hand, elongated magnetic elements, such as stripes and nanopillars,
can overcome energy barries by a nonuniform rotation of the magnetization,
for instance, via the motion of a domain wall across them. In this
cace the energy barrier depends on the cross-section but not on the
length of the magnetic elements.

The so-called micromagnetic approach \cite{bro63mic}, considering
magnetic materials within the continuous approximation at $T=0$ and
implemented in ready-to-use software packages, is hardly suitable
for these purposes. With the increasing of computer power, atomistic
approach considering magnetic materials as a collection of spins on
the lattice becomes more realistic. Within this framework, one can
accurately describe thermal properties such as the heat capacity,
temperature dependence of the aisotropy constants, and even phase
transitions. For most applications, spins can be considered as classical
and obeying the Landau-Lifshitz equation \cite{lanlif35} with the
Langevin stochastic fields (introduced in magnetism by Brown \cite{bro63pr})
mimicking the influence of the thermal bath. The latter lead to thermally-activated
escape out of metastable states. 

Thermal stability of metastable states of single-domain (SD) magnetic
particles was mainly investigated considering them as single spins
\cite{cofetal98prl,kalcof12jap} and using the Fokker-Planck equation.
Early numerical work on single spins with the stochastic Landau-Lifshitz-Langevin
(LLL) equation was done in Refs. \cite{libcha93jap,pallaz98prb}.
However, there are important many-spin aspects of magnetic nanoparticles,
especially related to the surface anisotropy \cite{garkac03prl,yanetal07prb}
(see also Ref. \cite{kacgar04springer} for a review). Non-uniform
magnetization via nucleation in magnetic nanoparticles was studied
in Ref. \cite{hinnow98prb} by Monte Carlo. Solving the LLL equation
for a collection of spins requires much more computing power and is
much more difficult \cite{chubykaloetal02prb,chunowchagar06prb,suhetal08prb,basatxchukac12prb,evansetal14jpc}.
One of the obstacles is the stochastic nature of these ordinary differential
equations (ODE) that forced using the second-order Heun ODE solver
with a rather short time step. Another difficulty is the need of generating
stochastic fields for the whole spin system at every integration step.
Recently the pulse-noise approach overcoming these two problems has
been proposed \cite{gar17pre}. The stochastic fields are represented
by rare pulses rotating spins at different angles around different
axes, and in the intervals between the pulses high-order ODE solvers
with larger time step can be used. The pulse-noise model is working
well in the cases of low and intermediate damping and at low temperatures,
where Langevin fields are weak.

In this paper, the pulse-noise approach is applied to uniform thermal
activation of single-domain magnetic particles, including a comparison
with the single-spin model, and to non-uniform thermal activation
via moving domain walls in elongated objects such as magnetic chains.
For this purpose, the crossover from uniform to nonuniform thermal
escape is studied analytically and formulas for the escape barrier
in different cases are worked out. This system was considered in Ref.
\cite{hinnow00prb} with the help of Monte Carlo and the LLL equation
at high damping. Here the focus is on the much less studied low damping
regime. While computations yield Arrhenius escape rates with correct
barriers in most cases, comparison with available theoretical results
for the prefactors is much more difficult. In particular, the thermal-activation
prefactor for SD particles is by more than order of magnitude higher
than that for the equivalent single spins.

The structure of the main part of the paper is the following. Section
\ref{sec:The-model} introduces the stochastic model of classical
spins, presents known results for thermal activation of single spins,
and reviews the pulse-noise method. Section \ref{sec:thermal-switching-SD}
shows the numerical results for single-domain magnetic particles in
comparizon with the single-spin model. In Sec. \ref{sec:Non-uniform-energy-barriers}
analytical results for elongated particles with transverse and longitudinal
fields are derived. Section \ref{sec:Thermal-switching-1D} investigates
the dynamics of thermal switching of magnetic chains in different
regimes and shows the numerical results for the escape rate $\Gamma(T)$
testing the barrier formulas of the preceding section. In Sec. \ref{sec:Linear-instability}
the linear-stability analysis of single-domain barrier states within
the chosen model is done. In Sec. \ref{sec:Variational-solution}
crossover between uniform and nonuniform thermal activation is studied
with the help of the variational method.

\section{The model and the methods\label{sec:The-model}}

Consider a parallelepiped-shape magnetic particle with a simple cubic
lattice described by the classical-spin Hamiltonian 
\begin{equation}
\mathcal{H}=\sum_{i}\left(-Ds_{zi}^{2}+D_{y}s_{yi}^{2}\right)-\mu_{0}\mathbf{H}\sum_{i}\mathbf{s}_{i}-\frac{1}{2}\sum_{ij}J_{ij}\mathbf{s}_{i}\cdot\mathbf{s}_{j},\label{Ham}
\end{equation}
where $|\mathbf{s}_{i}|=1,$ $\mathbf{H}$ is the magnetic field,
$\mu_{0}$ is the magnetic moment of the atom, $J_{ij}$ is the exchange
with $J$ being the nearest-neighbors exchange coupling, $D>0$ is
the uniaxial-anisotropy constant, and $D_{y}>0$ is the hard-axis
$y$ anisotropy. One can also include a surface anisotropy and the
dipole-dipole interaction. The latter, for single-domain particles,
gives rise to the biaxial anisotropy, $D_{y}>0$, if the particle
has a shape flat in the $y$ direction.

The dynamics is described by the Landau-Lifshitz-Langevin (LLL) equation
\begin{equation}
\mathbf{\dot{s}}_{i}=\gamma\left[\mathbf{s}_{i}\times\left(\mathbf{H}_{\mathrm{eff},i}+\boldsymbol{\zeta}_{i}\right)\right]-\gamma\lambda\left[\mathbf{s}_{i}\times\left[\mathbf{s}_{i}\times\mathbf{H}_{\mathrm{eff},i}\right]\right],\label{LLL}
\end{equation}
where $\mathbf{H}_{\mathrm{eff},i}\equiv-\partial\mathcal{H}/\partial\mathbf{s}_{i}$
is the effective field,
\begin{equation}
\mu_{0}\mathbf{H}_{\mathrm{eff},i}=\mu_{0}\mathbf{H}+2Ds_{zi}\mathbf{e}_{z}-2D_{y}s_{yi}\mathbf{e}_{y}+\sum_{j}J_{ij}\mathbf{s}_{j},
\end{equation}
 $\gamma$ is the gyromagnetic ratio, $\lambda$ is the dimensionless
damping constant \cite{lanlif35}, and $\boldsymbol{\zeta}_{i}$ are
stochastic field. Landau and Lifshitz have written the double-vector-product
relaxation term on general grounds. Later it was demonstrated that
the vector product in the noise term dictates the double-vector product
form of the damping term \cite{garishpan90tmf}. The stochastic model
above is equivalent to the Fokker-Planck equation introduced by Brown
for superparamagnetic particles \cite{bro63pr} (see also Ref. \cite{cofkal04book}).
The equilibrium solution of the Fokker-Planck equation should be a
Boltzmann distribution, that requires a relation between damping and
noise,
\begin{equation}
\left\langle \zeta_{\alpha,i}(t)\zeta_{\beta,j}(t')\right\rangle =\frac{2\lambda T}{\gamma\mu_{0}}\delta_{ij}\delta_{\alpha\beta}\delta(t-t'),
\end{equation}
where $k_{B}=1$ is set. Microscopic theories suggest $\lambda\ll1$.
In the case of spin magnetism with one type of magnetic atoms, one
has $\mu_{0}=g\mu_{B}S$, where $g=2$, $\mu_{B}$ is Bohr's magneton,
$S$ is the spin value of the magnetic atom, and $\gamma$ is given
by $\gamma=g\mu_{B}/\hbar$, so that $\gamma\mu_{0}=\left(g\mu_{B}\right)^{2}S/\hbar$.

In computations, $J$ and most other parameters and physical constants
are set to 1, so that one needs a relation between the time $t$ in
computations and the real time $t_{\mathrm{real}}$. For the original
system the exchange frequency is $\omega_{ex}=S^{2}J/\mu_{0}=SJ/\hbar$.
As this frequency is set to 1, the relation between the times reads
$t_{\mathrm{real}}=\frac{\hbar}{SJ}t$. For metallic Co, $J=93$ K
and $S=3/2$, so that $t_{\mathrm{real}}=5.3\times10^{-14}t$ s. The
maximal computation time in the present work $t=10^{6}$ corresponds
to $t_{\mathrm{real}}=5.3\times10^{-8}$ s or 53 ns for Co. It is
clear that computations cannot be extended to seconds and extrapolation
of the computational results is needed. Further, the uniaxial anisotropy
of Co is 0.22 K per atom, thus in $J$ units one has $D=0.22/96=0.0023$.
Pt alloys CoPt and FePt have a comparable exchange but a much stronger
anisotropy: $D=4.1/96=0.044$ for CoPt and $D=5.7/105=0.054$ for
FePt.

Uniaxial anisotropy creates an energy barrier $U$ between spin directions
parallel and antiparallel to $z$ axis. If the particle is in single-domain
(SD) state, that is realized for not too large sizes, than in the
absence of the field the barrier is given by $U=U_{SD}\equiv\mathcal{N}D$,
where $\mathcal{N}\equiv N_{x}N_{y}N_{z}$ is the number of spins
in the particle. Applied field lowers the barrier. The barrier exist
within the Stoner-Wohlfarth astroid \cite{stowoh48,stowoh91}, 
\begin{equation}
h_{x}^{2/3}+h_{z}^{2/3}\leq1,\qquad h_{x}\equiv\frac{\mu_{0}H_{x}}{2D},\qquad h_{z}\equiv\frac{\mu_{0}H_{z}}{2D}.\label{Sto-Woh}
\end{equation}
There are formulas for the barrier in the cases of purely transverse
and purely longitudinal field,
\begin{equation}
U_{SD}=\mathcal{N}D\left(1-h_{x}\right)^{2},\qquad U_{SD}=\mathcal{N}D\left(1-h_{z}\right)^{2}.\label{U_SD}
\end{equation}
For particle of a larger size, the barrier state becomes non-uniform
and for elongated particles it corresponds to a domain wall bisecting
the particle. In zero field, the non-uniform barrier is the energy
of a domain wall (DW) given by
\begin{equation}
U_{DW}=\mathcal{N}D\frac{4\delta}{L_{z}}=\frac{4\delta}{a}N_{x}N_{y}D,\qquad\delta=a\sqrt{\frac{J}{2D}},\label{U_DW}
\end{equation}
where $\delta$ is the domain-wall width, $a$ is the lattice spacing,
$L_{z}\equiv aN_{z}$ is the longest particle's dimension. The particle
is overcoming the barrier in the single-domain state if $U_{SD}<U_{DW}$
that amounts to $L_{z}\lesssim\delta$. The numerical coefficient
in this formula has to be worked out, including the cases with the
field applied.

At low temperatures, $T\ll U$, the particle is overcoming the barrier
via thermal activation that yields the escape rate of the Arrhenius
form
\begin{equation}
\Gamma=\Gamma_{0}e^{-U/T}.\label{Arrhenius}
\end{equation}
The process is described by Eq. (\ref{LLL}) or the equivalent Fokker-Planck
equation. The latter is suitable for the analytical work, especially
in the SD regime, and it allows to obtain the formula above with different
expressions for the prefactor $\Gamma_{0}$ in different parameter
regions. For a single spin, in the axially symmetric case $\Gamma_{0}\propto\lambda$.
If there is a saddle of the spin's energy (created by the transverse
field or transverse anisotropy), then there are three regimes of strong,
intermediate, and low damping. In the strong-damping case, $\Gamma_{0}\propto\lambda$,
if the Landau-Lifshitz equation is used. In the ID case $\Gamma_{0}$
is intependent of $\lambda$ that corresponds to the transition-state
theory. However, this regime is realized in a pretty narrow region
of $\lambda$. In the LD case $\Gamma_{0}\propto\lambda$ again. There
are numerous crossovers between these three regimes and the uniaxial
regime \cite{garkencrocof99pre}. Analytical solution of the multidimensional
Fokker-Planck equation for non-uniform thermal activation is a challenging
task. First, the barrier has to be found as the saddle point of the
particle's energy in a nonuniform state (see, e.g., \cite{braun90prl,braber94jap}).
The corresponding analytical results will be summarized below. Second,
calculation of the prefactor in Eq. (\ref{Arrhenius}) requires application
of functional methods and is especially nontrivial \cite{braun94prb}.

Numerically, the Fokker-Planck equation can be solved by the matrix-continued-fraction
method \cite{kalmykov00prb,kalcof12jap} that is very fast and accurate.
However, setting up equations for different kinds of anisotropy reqires
a serious work. For many-spin systems this method becomes unusable.

The most straightforward method of numerically solving the problem
of thermal activation (apart of the time-quantified Monte Carlo \cite{nowchaken00prl,chubykaloetal03prb})
is using Eq. (\ref{LLL}). The particle is prepared in a collinear
state corresponding to one of the energy minima, then Eq. (\ref{LLL})
is solved until the particle crosses the barrier. This yields the
first-passage time for a given computation. Such computations can
be run in parallel, and the inverse of the mean first-passage time
is identified with the escape rate $\Gamma$. In fact, there is a
more efficient method of data processing described in the Appendix. 

\begin{figure}
\begin{centering}
\includegraphics[width=9cm]{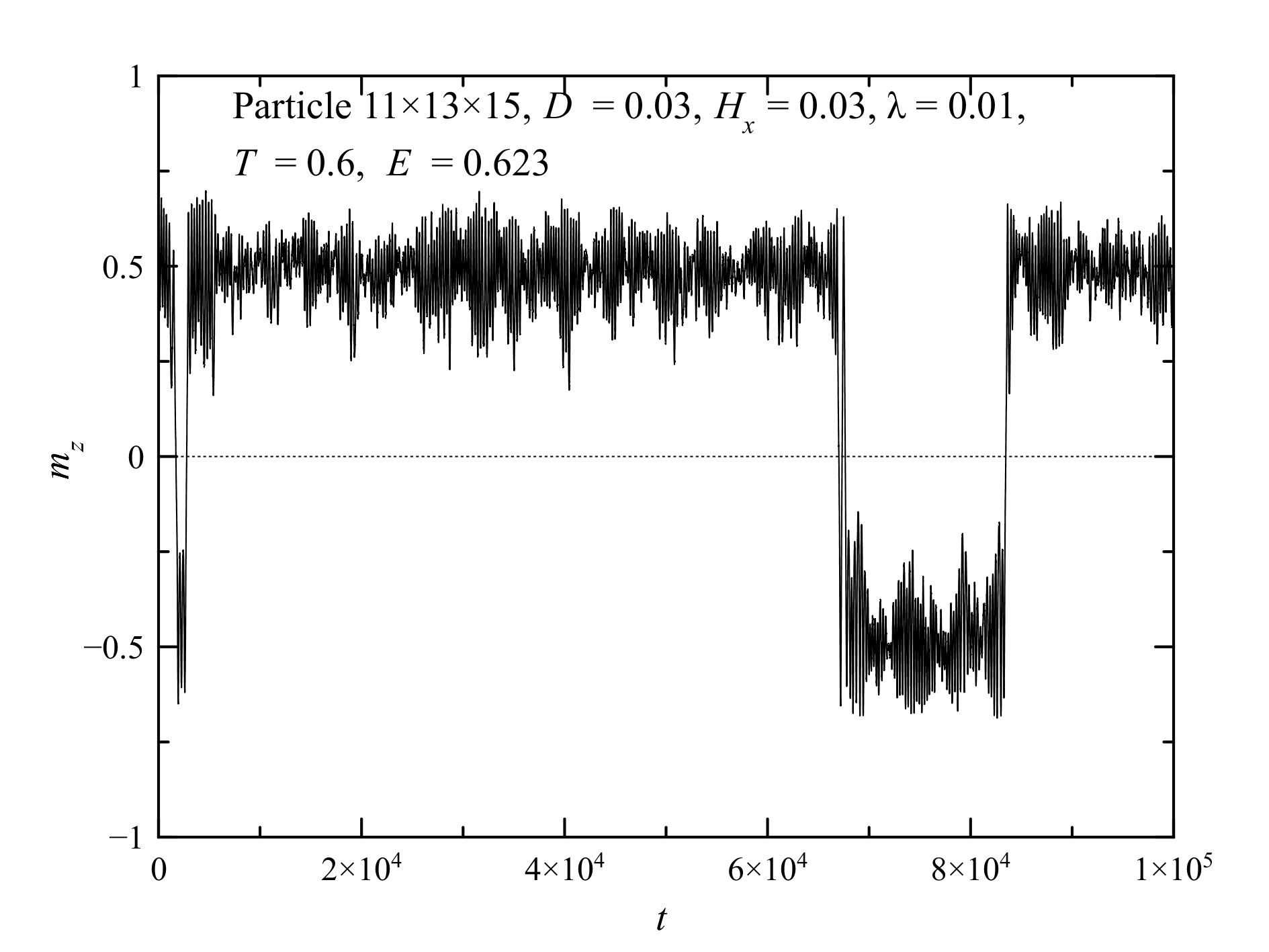}
\par\end{centering}
\begin{centering}
\includegraphics[width=9cm]{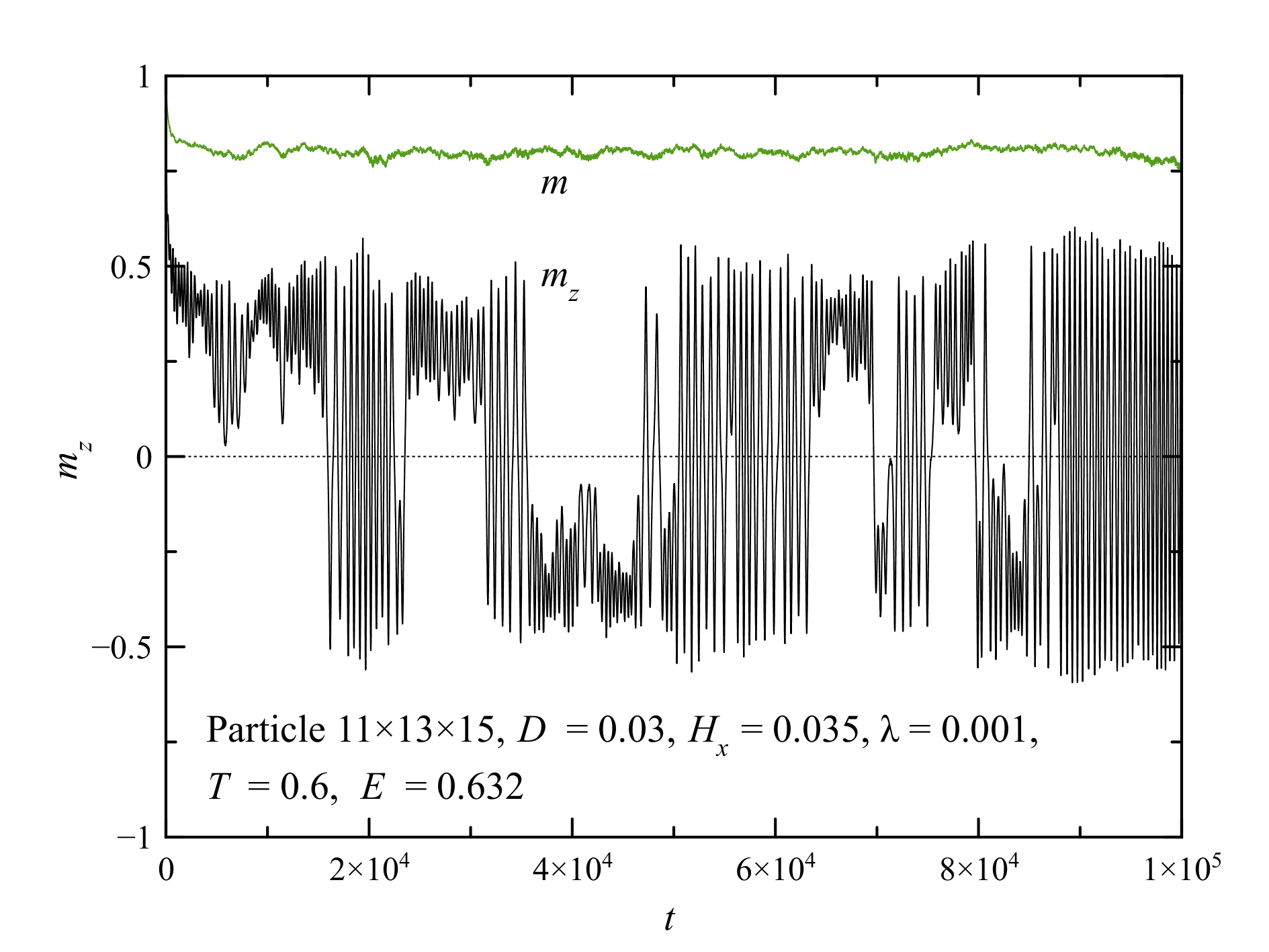}
\par\end{centering}
\caption{Thermal switching dynamics of a single-domain particle with transverse
field at elevated temperature $T=0.6$ for different values of the
damping $\lambda$. Upper panel: $\lambda=0.01$; Lower panel: $\lambda=0.001$
(energy diffision regime).}

\label{Fig_mz_vs_t_L=00003D11x13x15_Dz=00003D0.03_Hx=00003D0.03_T=00003D0.6}
\end{figure}

At low temperatures where $\Gamma$ is exponentially small, the computations
leading to particle's escape are very long. Usually the second-order
Heun method with a rather short time step is used to integrate the
stochastic LLL equation. For this reason, the LLL equation was mainly
solved for single-domain thermal activation considering the magnetic
particle as a single spin.

It was recently shown that in the relevant regions of intermediate
and low damping the continuous noise can be replaced by a pulse noise
with time interval $\Delta t$ that has to satisfy the two conditions
\begin{equation}
\varLambda_{N}\varDelta t\ll1,\qquad\gamma\lambda H_{\mathrm{eff}}\Delta t\ll1,
\end{equation}
where $\varLambda_{N}\equiv2\gamma\lambda T/\mu_{0}$ is the so-called
Néel attempt frequency, i.e., the high-temperature relaxation rate.
The pulse consists in rotating spins by the angles
\begin{equation}
\boldsymbol{\varphi}_{i}=\sqrt{\varLambda_{N}\varDelta t}\mathbf{G}_{i},\label{eq:phi_n}
\end{equation}
where $\mathbf{G}_{i}$ is a realization of a three-component vector,
each component being a normal distribution with a unit dispersion.
In the interval between the pulses, high-order numerical integrators
(for instance, the classical RK4 or Butcher's RK5, see, e.g., the
Appendix of Ref. \cite{gar17pre}) for the damped equations without
noise can be used. The step $\delta t$ of the numerical integration
can be chosen much greater than that used with the Heun method, that
gives a considerable speed-up. At low damping $\lambda$, one can
choose $\delta t\ll\varDelta t$ that makes the contribution of the
random-number generation into the computation time negligible. Thus
the speed of the method becomes the same as the speed of noiseless
computations. This allows to numerically solve the problem of thermal
activation of magnetic particles within the many-spin model, including
the case of non-uniform thermal activation. In the sequel, some of
such problems will be considered.

Typically, numerical solution used $\varDelta t=1$ and $\delta t=0.1$
with Butcher's RK5 ODE solver. It should be noted that, although at
low temperatures the dynamics is governed by $D$ and $H$ that define
the precession frequency of the particle, the required integration
step $\delta t$ is dictated by the exchange $J$. If $J\delta t\ll1$
is not satisfied, explicit integrators used here show instabilities.
The latter usually happen for $J\delta t>0.25.$ The physical reason
is that for large $J$ increasing the spin noncollinearity with the
neighbors leads to a very fast precession that a large-step integrator
cannot handle. This aspect is absent in the one-spin models of magnetic
particles and it makes a problem for a small ratio $D/J$. 

The main computer used in these computations was Dell Precision T7610
Workstation with two Dual Intel\textregistered{} Xeon\textregistered{}
Processors E5-2680, each having 10 cores. The algorithm was implemented
within Wolfram Mathematica with compilation and parallelization. In
computations, the energy was measured in the $J$ units, that is,
$J$ was set to 1, together with $\mu_{0}$, $\gamma$, and the lattice
spacing $a$. The bulk of computations was performed on particles
with free boundary conditions.

\section{thermal switching of single-domain magnetic particles\label{sec:thermal-switching-SD}}

\begin{figure}
\begin{centering}
\includegraphics[width=9cm]{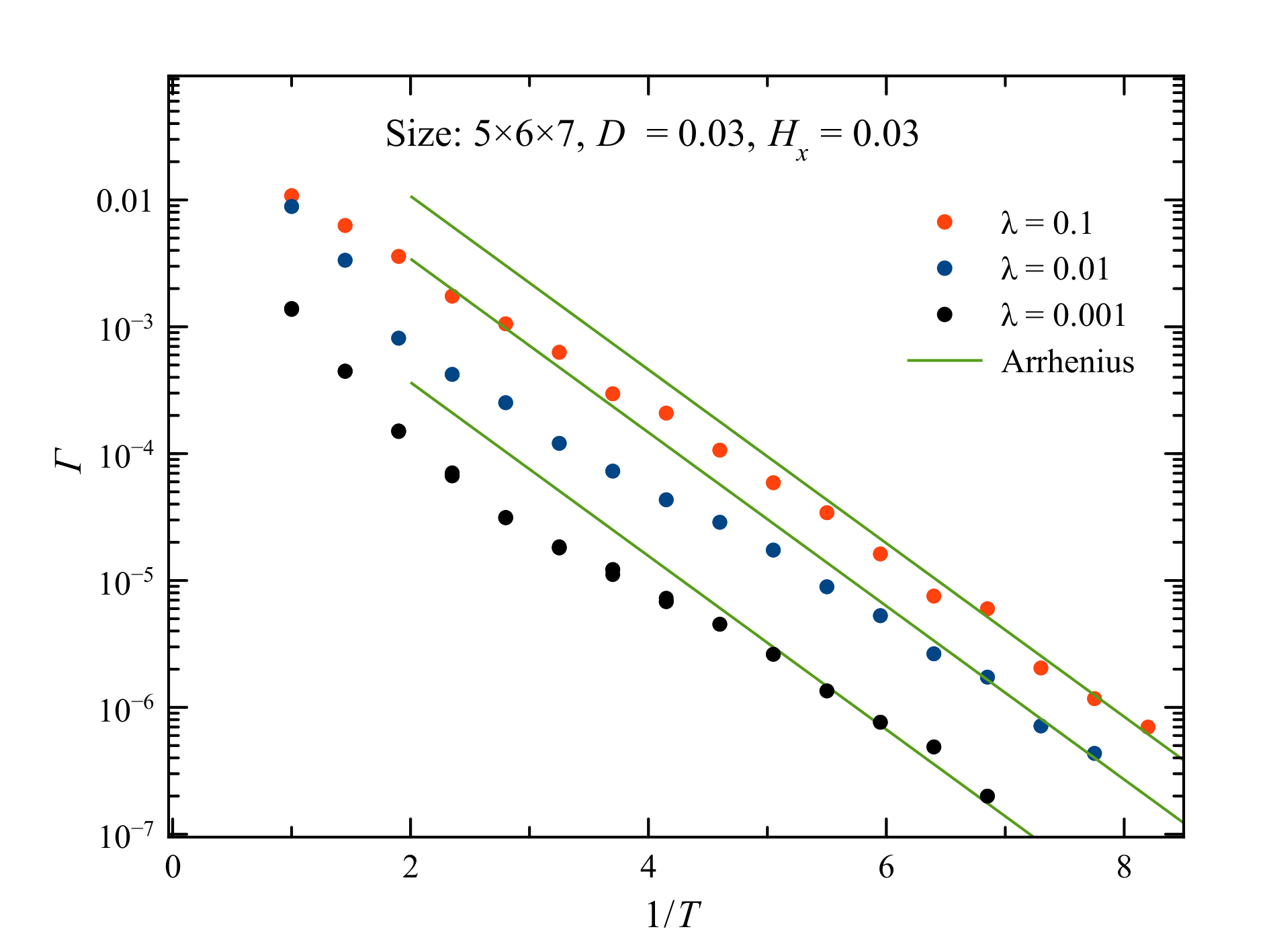}
\par\end{centering}
\begin{centering}
\includegraphics[width=9cm]{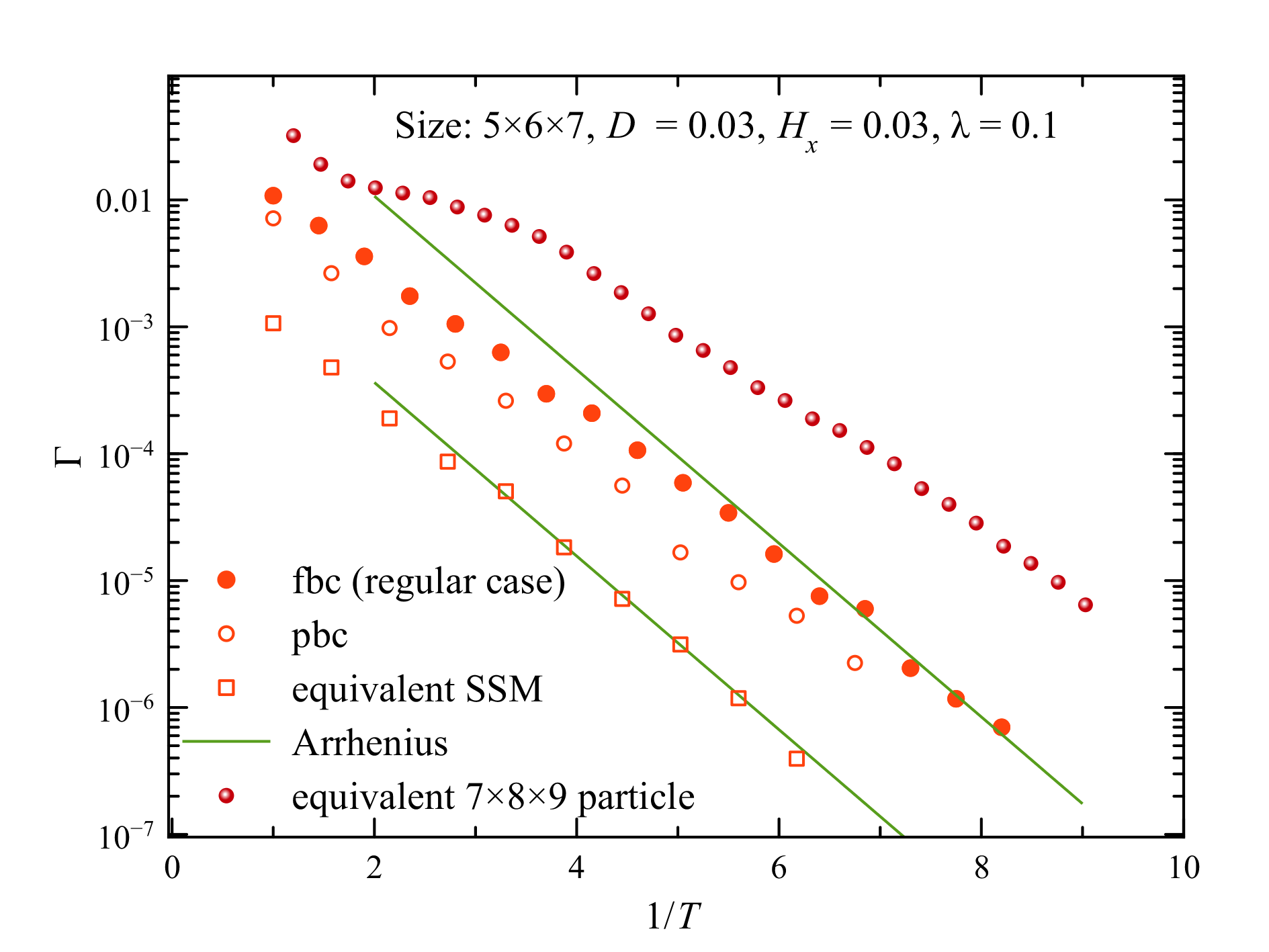}
\par\end{centering}
\caption{Thermal escape rate for the $5\times6\times7$ particle with $D=H_{x}=0.03$.
Upper panel: different values of the damping constant $\lambda=0.1$,
0.01, and 0.001. Lower panel: Systems with free and periodic boundary
conditions, compared with the results for an equivalent single spin
and for the equivalent $7\times8\times9$ particle.}
\label{Fig_Gam_vs_T_Nx=00003D5_Ny=00003D6_Nz=00003D7}
\end{figure}

In this section, thermal activation of particles in the single-domain
regime was studied. Fig. \ref{Fig_mz_vs_t_L=00003D11x13x15_Dz=00003D0.03_Hx=00003D0.03_T=00003D0.6}
shows the activation dynamics and transitions between the energy minima
for a $11\times13\times15$ particle containing 2145 spins with $D=0.03$
and transverse field at $T=0.6$ that is not small in comparison with
the bulk Curie temperature $T_{C}=1.444J$ of the classical 3D Heisenberg
model. In the upper panel, the results for the particle with $H_{x}=0.03$
and $\lambda=0.01$, prepared in the collinear state in the energy
minimum at $\theta=\arcsin h_{x}=30^{\circ}$ to $z$ axis, show robust
jumps between the energy minima characteristic to strong-to-intermediate
damping. The magnitude $m$ of the particle's magnetization 
\begin{equation}
\mathbf{m}\equiv\frac{1}{\mathcal{N}}\sum_{i}\mathbf{s}_{i}\label{m-Def}
\end{equation}
demonstrates a considerable thermal disordering that one can see from
the value $m\approx0.8$ in the lower panel. The average energy of
the system $E$, with respect to that of the fully ordered state,
is close to $T$. In the lower panel, dynamics of the particle with
$\lambda=0.001$ is clearly underdamped. This is the energy-diffusion
regime, in which, as particle acquires an energy above the barrier,
it begins crossing it repeatedly. $m$ is not changing essentially
during crossing the barrier in the above computations, thus one can
conclude that the particle remains in the single-domain state.

Single-domain particles are usually thought of as effective single
spins. Of course, the single-spin model (SSM) is much easier for computations
than the original many-spin model. The corresponding mapping can be
obtained by setting up the equation of motion for $\mathbf{m}$ of
Eq. (\ref{m-Def}) using Eq. (\ref{LLL}) for tightly bound spins.
The result has the same form as the latter, however, with the global
Langevin field
\begin{equation}
\boldsymbol{\Xi}=\frac{1}{\mathcal{N}}\sum_{i}\boldsymbol{\zeta}_{i}\label{Zeta_Def}
\end{equation}
whose correlators are given by
\begin{equation}
\left\langle \Xi_{\alpha}(t)\Xi_{\beta}(t')\right\rangle =\frac{2\lambda T}{\hbar\mathcal{N}}\delta_{\alpha\beta}\delta(t-t').
\end{equation}
Thus one can map the SD particle onto the single spin by introducing
the scaled temperature for the SSM
\begin{equation}
T_{\mathrm{SSM}}=T/\mathcal{N}.\label{OSM_mapping}
\end{equation}
 As the results for the single spin at the temperature $T_{\mathrm{SSM}}$
are expected be the same as the results for the SD particle at $T$,
one can plot the single spin results obtained at the temperature $T_{\mathrm{SSM}}$
vs $T=\mathcal{N}T_{\mathrm{SSM}}$ to compare with those for the
SD particle at $T$. 

To systematically study the temperature dependence of the thermal
activation rate in the SD regime, a $5\times6\times7$ particle containing
210 spins was used. Here, again, $D=H_{x}=0.03$ that corresponds
to $h_{x}=1/2$. The barrier value given by Eq. (\ref{U_SD}) is $U=210\times0.03\times0.25=1.575$.
Crossing the barrier was detected as the change of the sign of the
total magnetization projection $m_{z}$. The results for three different
values $\lambda=0.1$, 0.01, and 0.001 are shown in the upper panel
of Fig. \ref{Fig_Gam_vs_T_Nx=00003D5_Ny=00003D6_Nz=00003D7}. For
$T\lesssim0.2$ that corresponds to $U/T\gtrsim8$, the escape rate
follows the predicted Arrhenius behavior with the barrier value given
above. 

The lower panel of Fig. \ref{Fig_Gam_vs_T_Nx=00003D5_Ny=00003D6_Nz=00003D7}
shows the escape rates $\Gamma(T)$ for the $5\times6\times7$ particle
with free boundary conditions (the regular case), periodic boundary
conditions, and for the equivalent single-spin model. Although the
slope of all three curves is the same that indicates the same barrier,
the prefactors are essentially different: the highest prefactor for
the particle with fbc and the lowest one for the equivalent single
spin. 

In the figure also are shown the results for the larger $7\times8\times9$
particle containing 504 spins. The temperature for this particle in
the plot is scaled in the same way, i.e., the results are plotted
vs the scaled temperature $T=(210/504)T_{7\times8\times9}$, where
$T_{7\times8\times9}$ is the actual temperature of computations for
this model. One can see that here the deviation from the SSM is even
more pronouced: the prefactor is much larger and even the barrier
is noticeably lower. 

The effect of higher thermal escape rates for a many-spin particle
was observed earlier \cite{hinnow00prb} and attributed to thermal
disordering of the particle. Indeed, the temperature dependence of
the magnetization $m(T)$ for classical spin systems at low temperatures
is linear: $m(T)\cong1-cT/J.$ Here the coefficient $c$ depends on
the particle's size and on the boundary conditions at the surface
\cite{kacgar01pasurf}. For fbc there is an additional thermal disordering
at the surface, so that $c$ is higher than for the model with pbc.
For the SSM this effect is absent, $c=0$. For the magnetic particle,
the barrier acquires a linear temperature dependence via the effective
temperature-dependent anisotropy constant,
\begin{equation}
U(T)\cong U_{0}-bT,\qquad b\sim\mathcal{N}Dc/J.
\end{equation}
As the result, the prefactor in Eq. (\ref{Arrhenius}) increases by
the factor $\exp\left(b\right)$ that can be large, as is the case
here. The lower apparent barrier for the $7\times8\times9$ particle
must be a consequence of the higher temperature $T_{7\times8\times9}$
for which the magnetization decreases stronger than linearly. To summarize,
the mapping of the SD magnetic particle on the single-spin model is
incomplete, as the assumption of tightly-bound spins misses the important
effect of thermal disordering of the particle.

In quantum mechanics, there is the Bloch's law for the magnetization
at low temperatures, $m(T)=1-c'\left(T/J\right)^{3/2}$ and thus $U(T)=U_{0}-b'T^{3/2}$
with $b'\sim\mathcal{N}Dc'/J^{3/2}$. In this case the additional
temperature-dependent prefactor $\exp\left(b'\sqrt{T}\right)$ should
emerge. At the moment, however, it is unclear how to compute the thermally-activated
dynamics of quantum spins from first principles.

One can take into account the dynamics of the particle's magnetization
at elevated temperatures within the single-spin approach by using
the Landau-Lifshitz-Bloch (LLB) equation \cite{gar97prb} with added
longitudinal stochastic terms changing $m$ \cite{garfes04prb,evansetal12prb}.
Overcoming the barrier, the particle decreases its magnetization,
up to its compete disappearance in the barrier state at temperatures
close to the Curie temperature \cite{kacgar01pafree,kazhinchanow09epl}.
This should essentially change particle's dynamics and thus the thermal-activation
prefactor. Note a similar phenomena in the physics of domain walls:
in the temperature interval below the Curie temperature the structure
of the DW changes so that there is only $z$ component of the magnetization
that changes its sign going through zero \cite{bulgin64jetp} and
changing the domain-wall dynamics completely \cite{koegarharjah93prl,harkoegar95prb}.

\begin{figure}
\begin{centering}
\includegraphics[width=9cm]{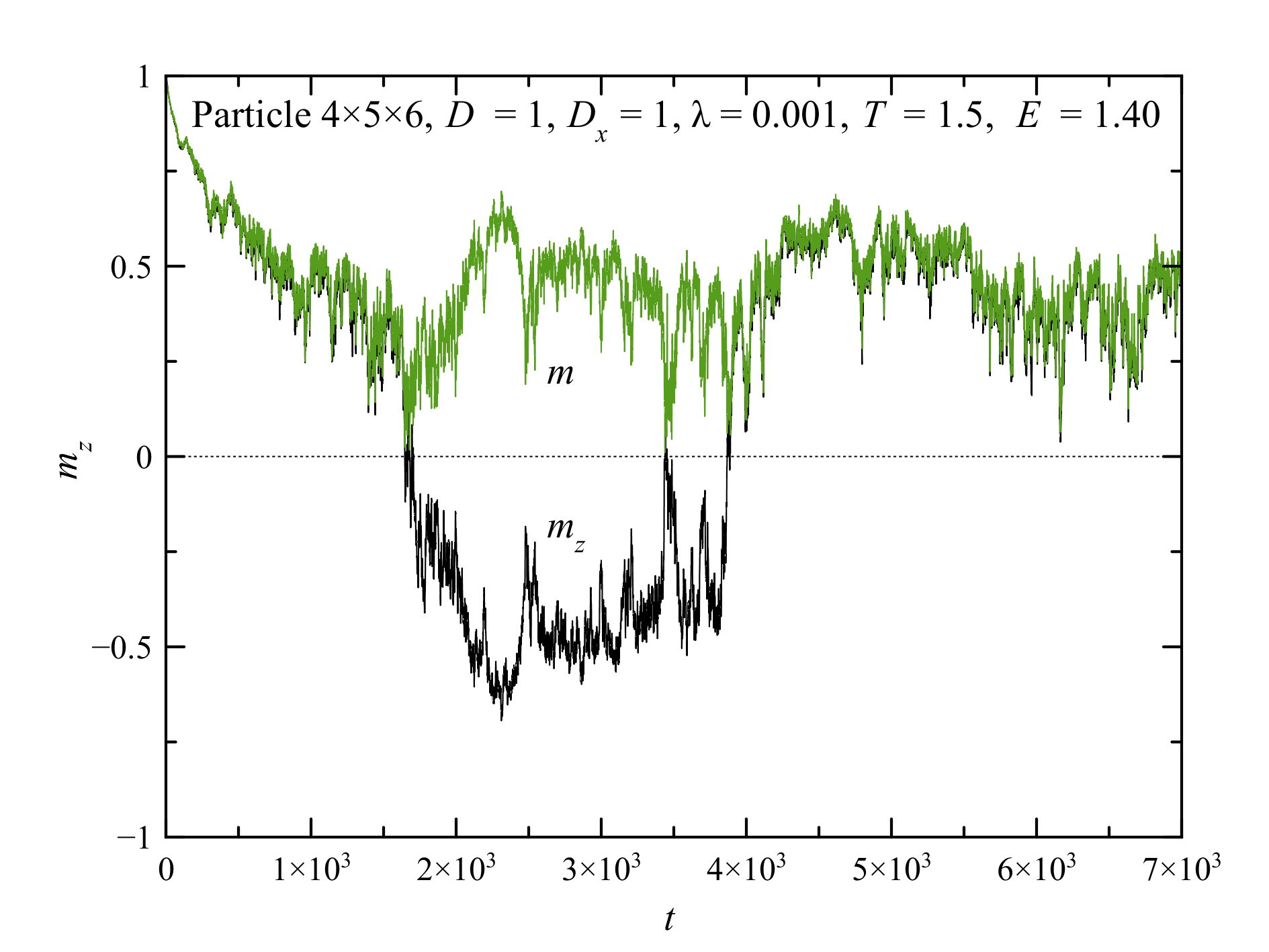}
\par\end{centering}
\begin{centering}
\includegraphics[width=9cm]{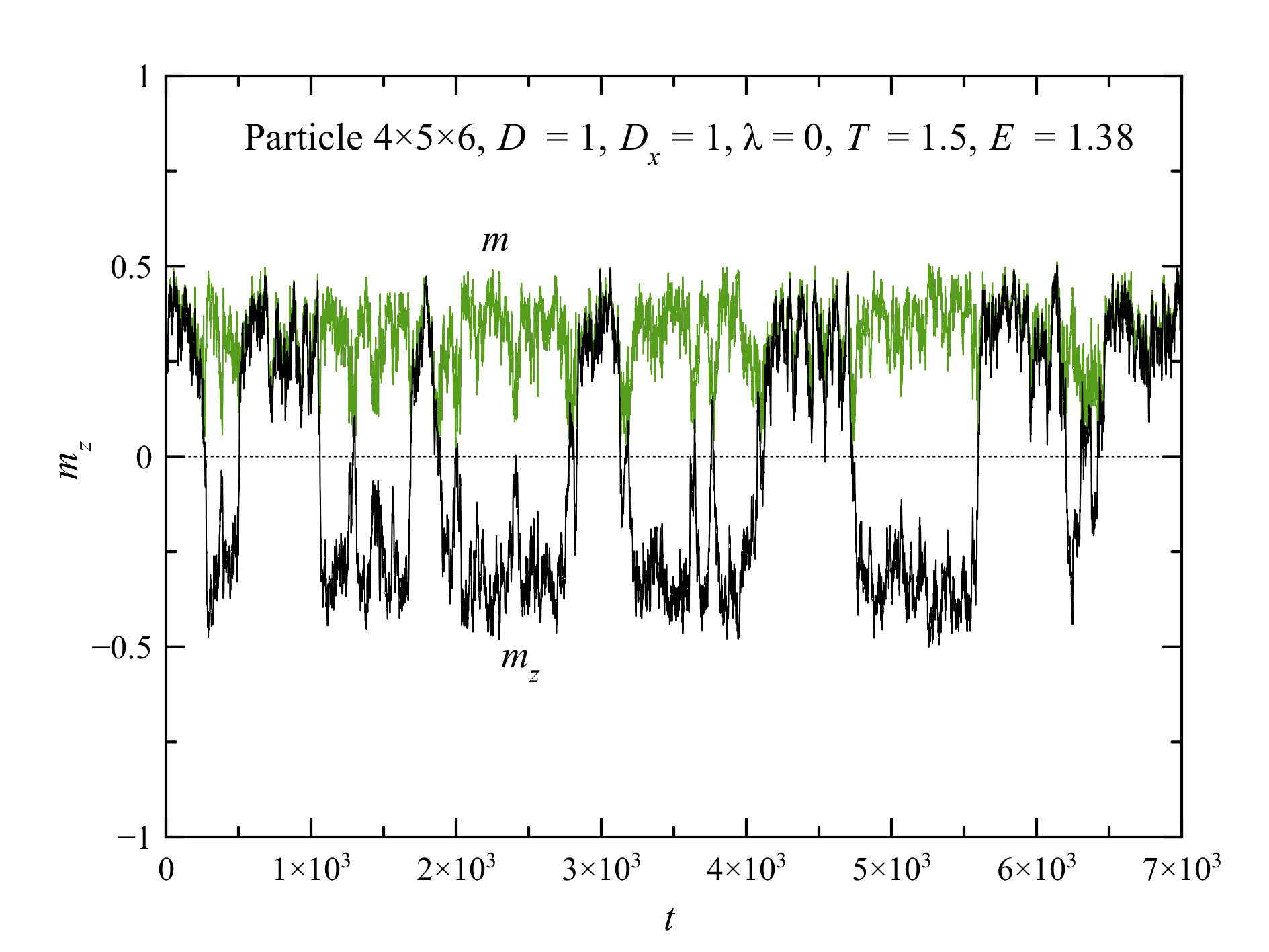}
\par\end{centering}
\caption{Thermal switching via changing the magnitude of the magnetization
$m$ of a $4\times5\times6$ particle with biaxial anisotropy $D=D_{x}=1$
at high temperature, $T=1.5$. Upper panel: $\lambda=0.001$; Lower
panel: $\lambda=0$.}

\label{Fig_mz_vs_t_L=00003D4x5x6_Dz=00003D1_Dx=00003D1_T=00003D1.5}
\end{figure}

In fact, these phenomena can be observed in the present computations
on the many-spin model for strong anisotropy and high temperature.
Fig. \ref{Fig_mz_vs_t_L=00003D4x5x6_Dz=00003D1_Dx=00003D1_T=00003D1.5}
shows thermal switching dynamics of a $4\times5\times6$ particle
of 120 spins with biaxial anisotropy $D=D_{x}=1$ at a high temperature,
$T=1.5$. In the upper panel, $\lambda=0.001$, first the particle
thermalizes starting from the collinear initial state that takes some
time because of the low damping, then it begins to jump between the
energy minima. One can see that $m_{z}$ correlates with $m$ and
that $m$ turns to zero when $m_{z}$ changes its sign. In the lower
panel, the computation was continued setting $\lambda=0$. The results
clearly show that the particle can cross the barrier even without
a coupling to the bath. The reason is that the magnetic particle has
internal degrees of freedom, spin waves, that can serve as the particle's
own bath. The contribution from internal spin waves enters the precessional
equation of motion for the particle's magnetization, Eq. (21) of Ref.
\cite{garkac09prb}, and it could play the role of thermal noise helping
to surmount the barrier. It is interesting that switching off the
damping increased the rate of overbarrier transitions. Although the
dynamics is conservative, the results look incoherent that is a consequence
of a strong thermal spin disordering in the cluster. Since the directions
of neighboring spins strongly deviate at such high temperature, it
is difficult to decide whether the magnetization switching is uniform
or not, especially in small clusters.

\begin{figure}
\begin{centering}
\includegraphics[width=9cm]{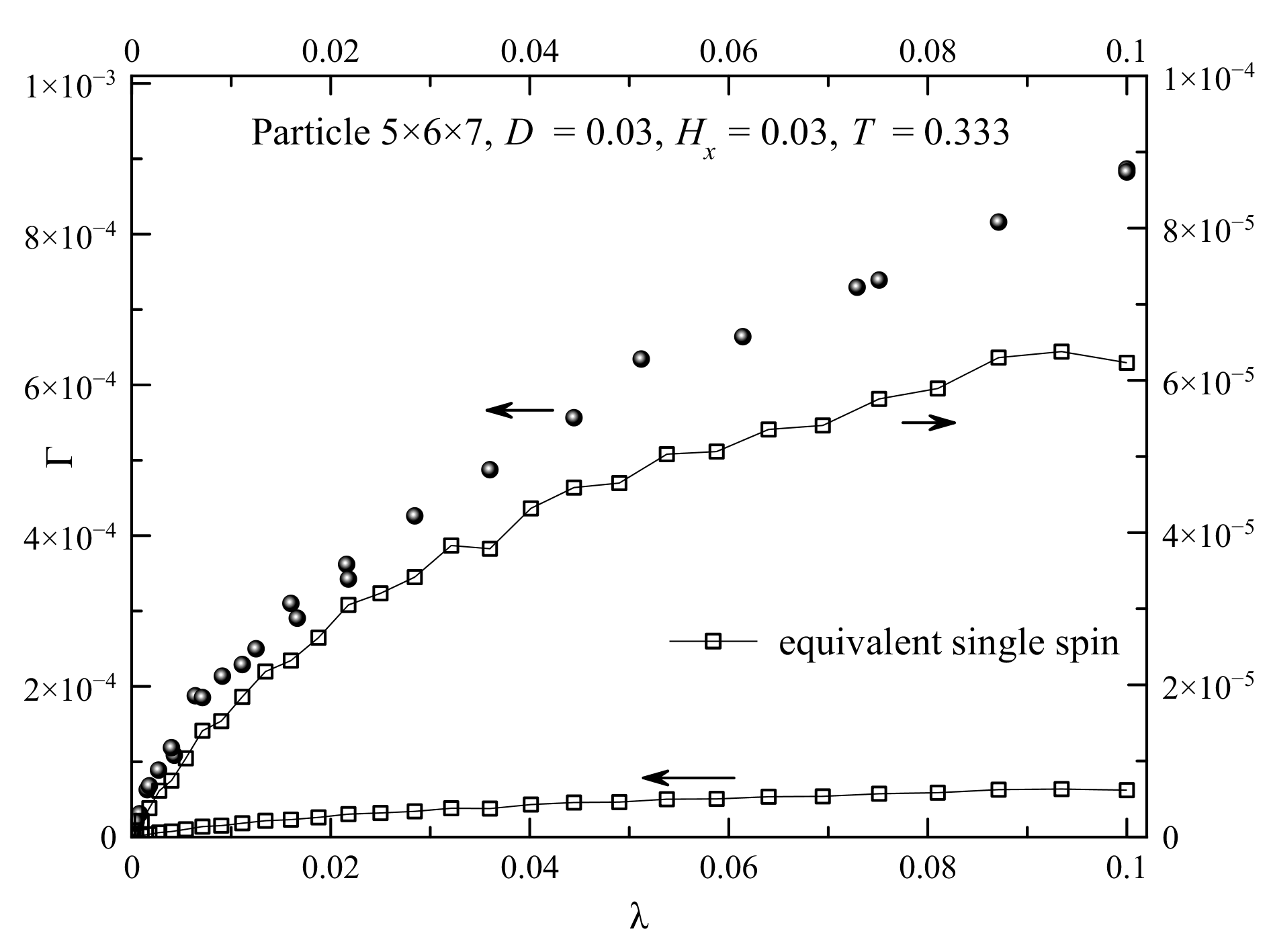}
\par\end{centering}
\begin{centering}
\includegraphics[width=9cm]{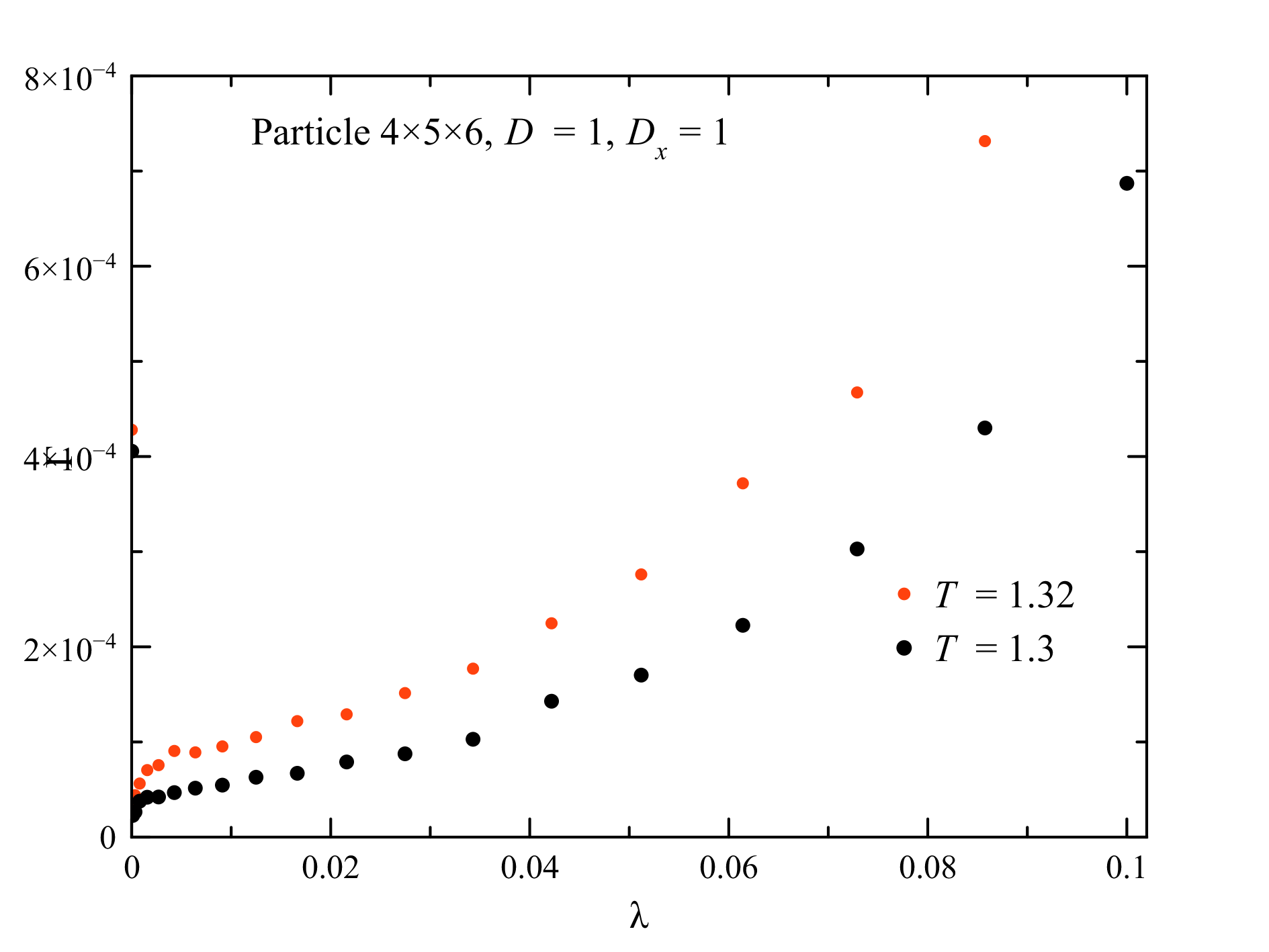}
\par\end{centering}
\caption{Thermal escape rate vs $\lambda$. Upper panel: $5\times6\times7$
particle with $D=H_{x}=0.03$ at $T=0.333$, compared with the result
for the equivalent single spin. For the latter, the escape rate is
by more than an order of magnitude lower. Lower panel: $4\times5\times6$
particle with $D=D_{x}=1$ at $T=1.3$. Thermal switching via longitudinal
relaxation, even in the absence of the coupling to the external bath.}

\label{Fig_Gam_vs_lam}
\end{figure}

The $\lambda$ dependence of the escape rate in the uppper panel of
Fig. \ref{Fig_Gam_vs_lam} shows mainly the low-damping regime with
a beginning of a crossover to the intermediate-damping regime at largest
$\lambda$. Escape rate vanishing at $\lambda\rightarrow0$ is expected
for a single spin. However, for a particle this is nontrivial since
the particle has its own internal bath. The likely reason is that
the contribution of the internal spin waves is quadratic in the anisotropy
that does not break the translational invariance \cite{garkac09prb},
such as the volume anisotropy in Eq. (\ref{Ham}), so that the effect
of the internal bath is weak for small anisotropies and low temperatures.
For strong anisotropy and high temperature, escape via the internal
bath at $\lambda=0$ is possible, as can be seen in the lower panel
of Fig. \ref{Fig_Gam_vs_lam}. The inverted curvature of the $\lambda$-dependence
of the switching rate indicates a different type of dynamics, likely
the longitudinal relaxation. 

It sould be noted that these computations start from the collinear
spin state and it requires a warming time to reach the preset temperature
$T$. In the limit $\lambda\rightarrow0$, the warming time goes to
infinity, this is why $\Gamma$ vanishes at the smallest $\lambda$
in the lower panel of Fig. \ref{Fig_Gam_vs_lam}. To study thermal
switching at $\lambda=0$, pre-warming is needed, as in the lower
panel of Fig. \ref{Fig_mz_vs_t_L=00003D4x5x6_Dz=00003D1_Dx=00003D1_T=00003D1.5}.

\section{Non-uniform energy barriers for long particles\label{sec:Non-uniform-energy-barriers}}

As said above, there are many different limits for the thermal activation
prefactor $\Gamma_{0}$, especially for the non-uniform thermal activation.
Thus here the analytical attention will be given to the barrier $U$,
while prefactors can be determined numerically.

To find the non-uniform barrier, one needs the continuous approximation
for the particle's energy, Eq. (\ref{Ham})
\begin{equation}
\mathcal{H}=\frac{1}{a^{3}}\intop dV\left\{ \frac{1}{2}a^{2}J\left(\frac{\partial s_{\alpha}}{\partial\mathbf{r}}\right)^{2}-Ds_{z}^{2}+D_{y}s_{y}^{2}-\mu_{0}\mathbf{H}\cdot\mathbf{s}\right\} \label{Ham_continuous}
\end{equation}
with summation over the repeated $\alpha$. Minimizing this energy
leads to the equation
\begin{equation}
\mathbf{s}\times\left(\mu_{0}\mathbf{H}+2Ds_{z}\mathbf{e}_{z}-2D_{y}s_{y}\mathbf{e}_{y}+a^{2}J\Delta\mathbf{s}\right)=0\label{equilibrium_eq}
\end{equation}
with the boundary condition 
\begin{equation}
\mathbf{s}\times\frac{\partial\mathbf{s}}{\partial r_{\alpha}}n_{\alpha}=0\label{Boundary_condition}
\end{equation}
at the surfaces, where $\mathbf{n}$ is the normal to the surface.

We will consider a particle elongated along $z$ axis and search for
the solution in the form $\mathbf{s}=(s_{x}(z),0,s_{z}(z))$, also
assuming $\mathbf{H}=(H_{x},0,H_{z})$. Then only $y$ component of
Eq. (\ref{equilibrium_eq}) is nonzero. In terms of the parameters
introduced in Eqs. (\ref{Sto-Woh}) and (\ref{U_DW}), one has
\begin{equation}
s_{z}\left(h_{x}+\delta^{2}s_{x}''\right)-s_{x}\left(h_{z}+s_{z}+\delta^{2}s_{z}''\right)=0,\label{DW_equation_1D}
\end{equation}
while the boundary condition becomes $s_{x}'=s_{z}'=0$. The energy,
Eq. (\ref{Ham_continuous}), in this case becomes
\begin{equation}
E=\frac{\mathcal{N}D}{L_{z}}\intop_{0}^{L_{z}}dz\left\{ \delta^{2}\left[\left(s_{x}'\right)^{2}+\left(s_{z}'\right)^{2}\right]+1-s_{z}^{2}-2\mathbf{h}\cdot\mathbf{s}\right\} .\label{E_prepared}
\end{equation}

In the transverse field, $h_{x}>0,$ the solution for the domain-wall
profile in the infinite system has the form 
\begin{equation}
s_{z}=\pm\sqrt{1-h_{x}^{2}}\frac{\sinh\frac{z-z_{0}}{\delta_{h}}}{h_{x}+\cosh\frac{z-z_{0}}{\delta_{h}}},\quad s_{x}=\frac{1+h_{x}\cosh\frac{z-z_{0}}{\delta_{h}}}{h_{x}+\cosh\frac{z-z_{0}}{\delta_{h}}},
\end{equation}
where $\delta_{h}\equiv\delta/\sqrt{1-h_{x}^{2}}$. The barrier state
corresponds to the DW in the center of the particle, $z_{0}=L_{z}/2$.
For $L_{z}\gg\delta$, the boundary conditions at the ends are practically
satisfied, and one can use this solution to calculate the barrier
energy using Eq. (\ref{E_prepared}) with $h_{z}=0$. The result has
the form
\begin{equation}
U_{DW}(h_{x})=4\mathcal{N}D\frac{\delta}{L_{z}}\left(\sqrt{1-h_{x}^{2}}-2h_{x}\arctan\sqrt{\frac{1-h_{x}}{1+h_{x}}}\right).\label{U_DW_hx}
\end{equation}
The limiting cases of this formula are
\begin{equation}
\frac{U_{DW}(h_{x})}{U_{DW}(0)}\cong\begin{cases}
1-\frac{\pi}{2}h_{x}, & h_{x}\ll1\\
\frac{2^{3/2}}{3}\left(1-h_{x}\right)^{3/2}, & 1-h_{x}\ll1.
\end{cases}
\end{equation}

In the longitudinal field, $h_{z}>0,$ one can use the saddle-point
solution found for the infinite system \cite{braun90prl}, in our
case centered at one of the particle's ends, say, near $z=0$,
\begin{equation}
\tan\frac{\theta}{2}=\sqrt{\frac{h_{z}}{1-h_{z}}}\cosh\left(\frac{\sqrt{1-h_{z}}}{\delta}z\right).
\end{equation}
The spin components are given by
\begin{equation}
s_{z}=\frac{1-\tan^{2}\left(\theta/2\right)}{1+\tan^{2}\left(\theta/2\right)},\qquad s_{x}=\frac{2\tan\left(\theta/2\right)}{1+\tan^{2}\left(\theta/2\right)},
\end{equation}
where the sign in front of $s_{z}$ is chosen so that $s_{z}=-1$
($\theta=\pi$) at infinity (the metastable state for $h_{z}>0$).
This solution satisfies Eq. (\ref{DW_equation_1D}) with $h_{x}=0$
and the boundary conditions. One has $s_{z}(0)=1-2h_{z}.$ For $h_{z}\ll1$
this yields $s_{z}(0)\cong1$, whereas the point at which $s_{z}=0$
($\theta=\pi/2,$ ) is far from the end of the particle. This is a
domain wall in the particle's bulk. For $1-h_{z}\ll1$ one has $s_{z}(0)\cong-1$
that is a very small deviation from the metastable state.

\begin{figure}
\begin{centering}
\includegraphics[width=9cm]{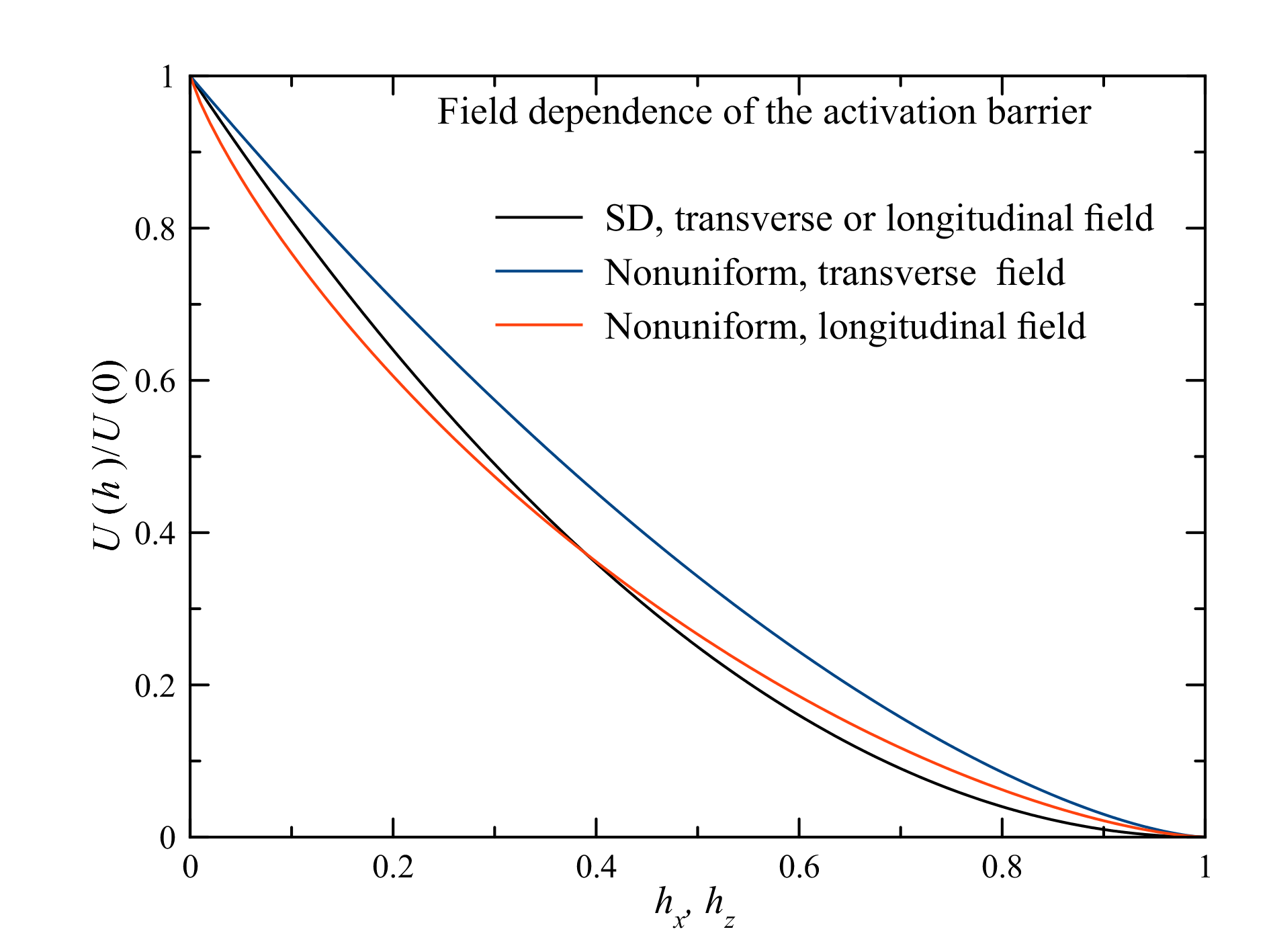}
\par\end{centering}
\caption{Field dependence of the energy barrier for magnetic particles in the
uniform (SD) and nonuniform regimes.}

\label{Fig_U_vs_h}
\end{figure}

The energy barrier is equal to the difference of the barrier energy
and the energy of the metastabe state $\propto2h_{z}$. Eq. (\ref{E_prepared})
with $h_{x}=0$ yields
\begin{equation}
U(h_{z})=4\mathcal{N}D\frac{\delta}{L_{z}}\left[\sqrt{1-h_{z}}-\frac{h_{z}}{2}\ln\frac{1+\sqrt{1-h_{z}}}{1-\sqrt{1-h_{z}}}\right].
\end{equation}
The limits of this formula are
\begin{equation}
\frac{U(h_{z})}{U_{DW}(0)}=\begin{cases}
1-\frac{h_{z}}{2}\left(1+\ln\frac{4}{h_{z}}\right), & h_{z}\ll1\\
\frac{2}{3}\left(1-h_{z}\right)^{3/2}, & 1-h_{z}\ll1.
\end{cases}
\end{equation}
Fields dependences of the barrier worked out above, as well as Eq.
(\ref{U_SD}) for the SD particle, are shown in Fig. \ref{Fig_U_vs_h}. 

\section{Thermal switching of spin chains\label{sec:Thermal-switching-1D}}

\begin{figure}
\begin{centering}
\includegraphics[width=9cm]{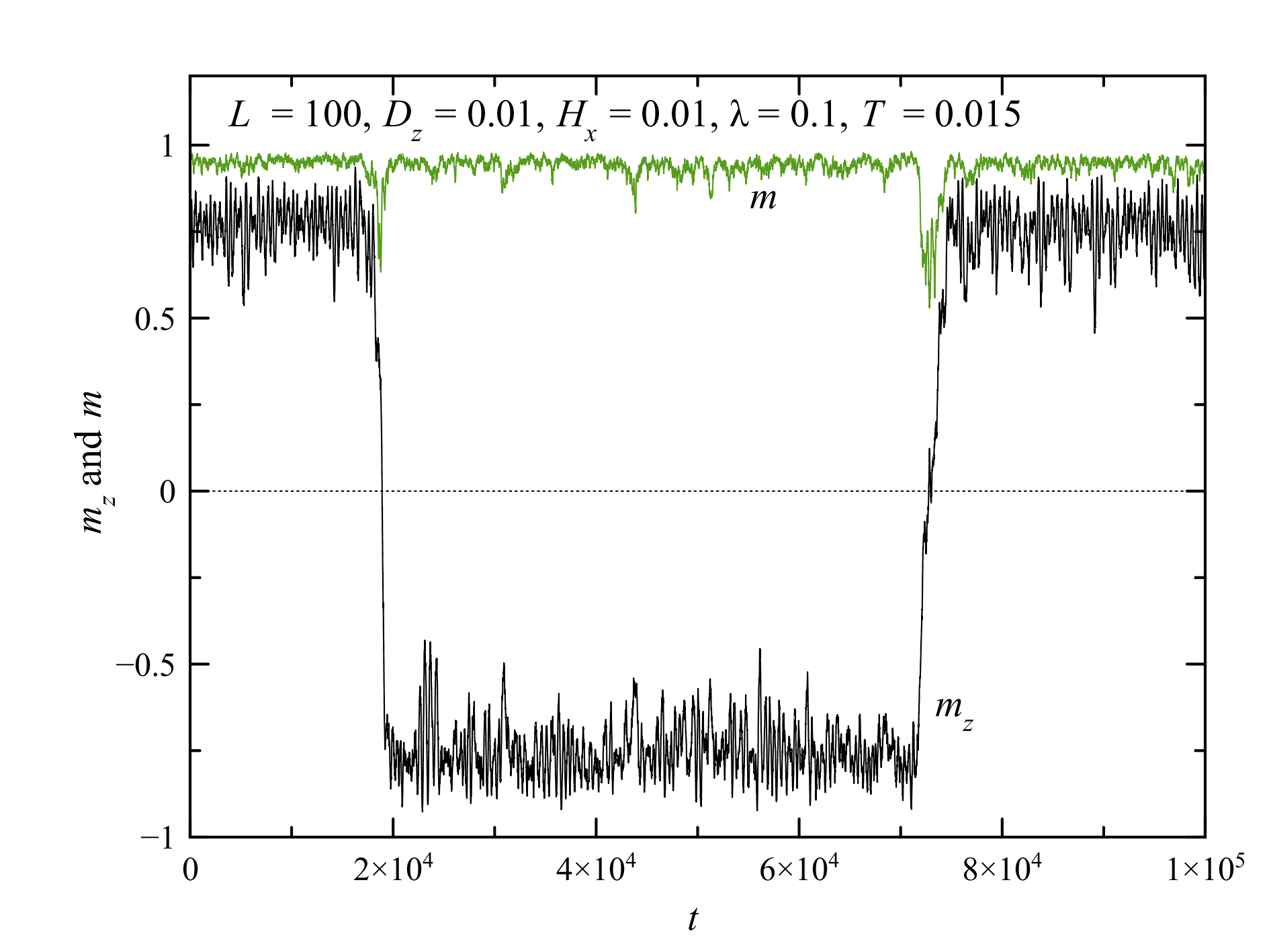}
\par\end{centering}
\begin{centering}
\includegraphics[width=9cm]{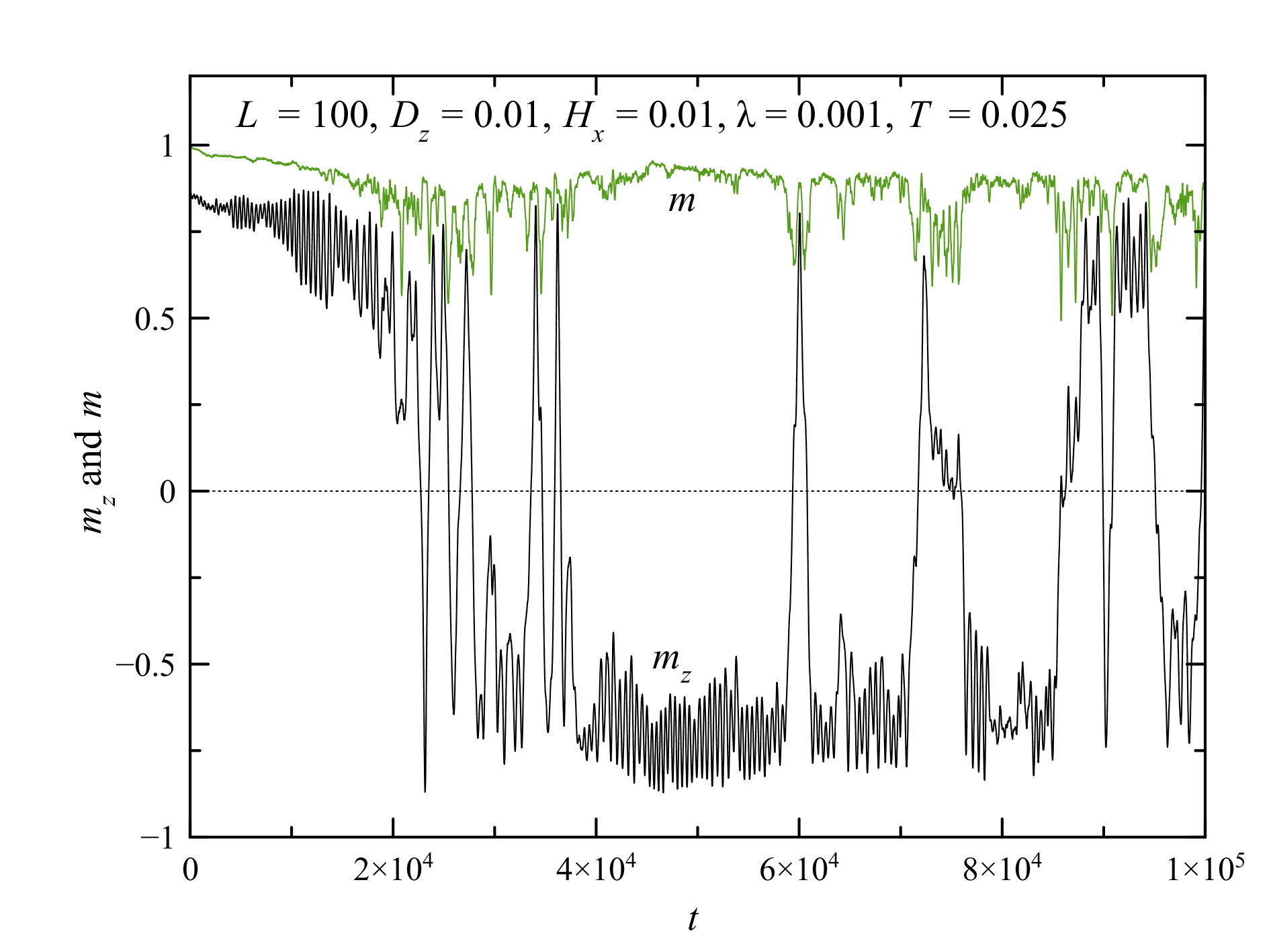}
\par\end{centering}
\caption{Thermal switching dynamics of a magnetic chain of $100$ spins with
transverse field for different values of the damping $\lambda$. Upper
panel: $\lambda=0.01$; Lower panel: $\lambda=0.001$ (energy diffision
regime). Unlike the SD particle in Fig. \ref{Fig_mz_vs_t_L=00003D11x13x15_Dz=00003D0.03_Hx=00003D0.03_T=00003D0.6},
here domain walls are traversing the chain in different directions,
changing the sign of $m_{z}$.}

\label{Fig_mz_vs_t_L=00003D100_Dz=00003D0.01_Hx=00003D0.01_T=00003D0.015}
\end{figure}
The simplest realization of elongated particles are spin chains. A
chain of $L=100$ particles with $D=0.01$ has the DW width $\delta\simeq7\ll L$
and is expected to overcome the barries at low temperatures via a
moving domain wall. The results for thermal switching dynamics for
this chain in the transverse field $H_{x}=0.01$ at $T=0.015$ and
0.025 for the intermediate and low damping $\lambda$, shown in Fig.
\ref{Fig_mz_vs_t_L=00003D100_Dz=00003D0.01_Hx=00003D0.01_T=00003D0.015},
are similar to those for the single-domain particle in Fig. \ref{Fig_mz_vs_t_L=00003D11x13x15_Dz=00003D0.03_Hx=00003D0.03_T=00003D0.6}.
The magnetization magnitude $m$ decreases when the chain is crossing
the barrier ($m_{z}=0$) but this decrease is much less than 50\%,
so that the barrier state of the chain is closer to a single-domain
state than to a state with a moving domain wall that would result
in $m=0$ when the DW is in the chain's center. Spin configurations
corresponding to crossing the barrier show spins nearly perpendicular
to $z$ axis with noticeable disordering due to thermal spin waves.
Observation of thermal activation via a moving domain wall in this
case requires lowering the temperature.

\begin{figure}
\begin{centering}
\includegraphics[width=9cm]{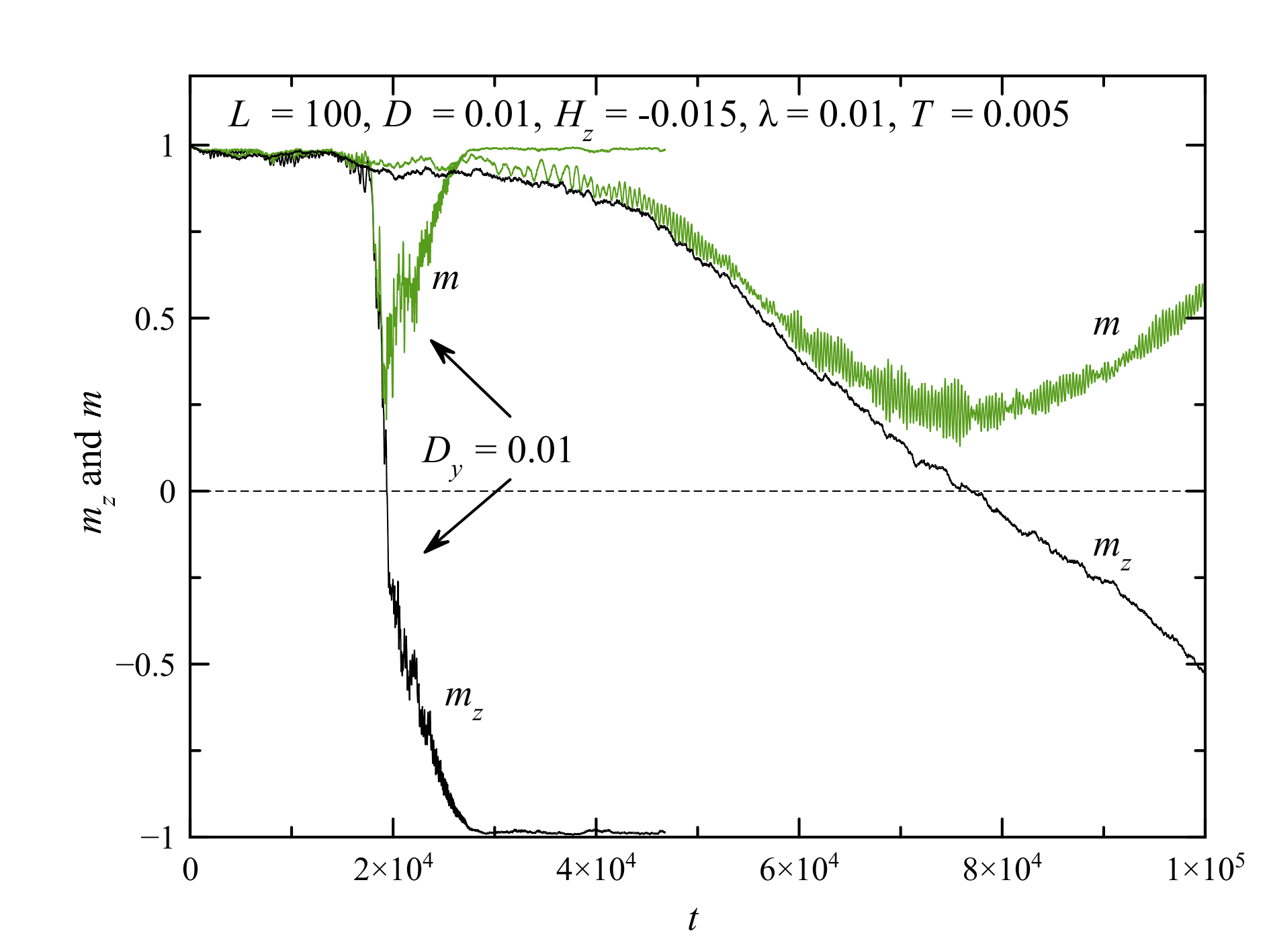}
\par\end{centering}
\caption{Thermal switching of a field-biased magnetic chain of $100$ spins
occurs via moving domain walls as the magnetization $m$ strongly
decreases in the process. Adding the hard-axis $y$ anisotropy $D_{y}$
speeds up the motion of the domain wall.}

\label{Fig_mz_vs_t_L=00003D100_Dz=00003D0.01_Dy=00003D0,0.01_Hz=00003D-0.015}
\end{figure}
\begin{figure}
\begin{centering}
\includegraphics[width=9cm]{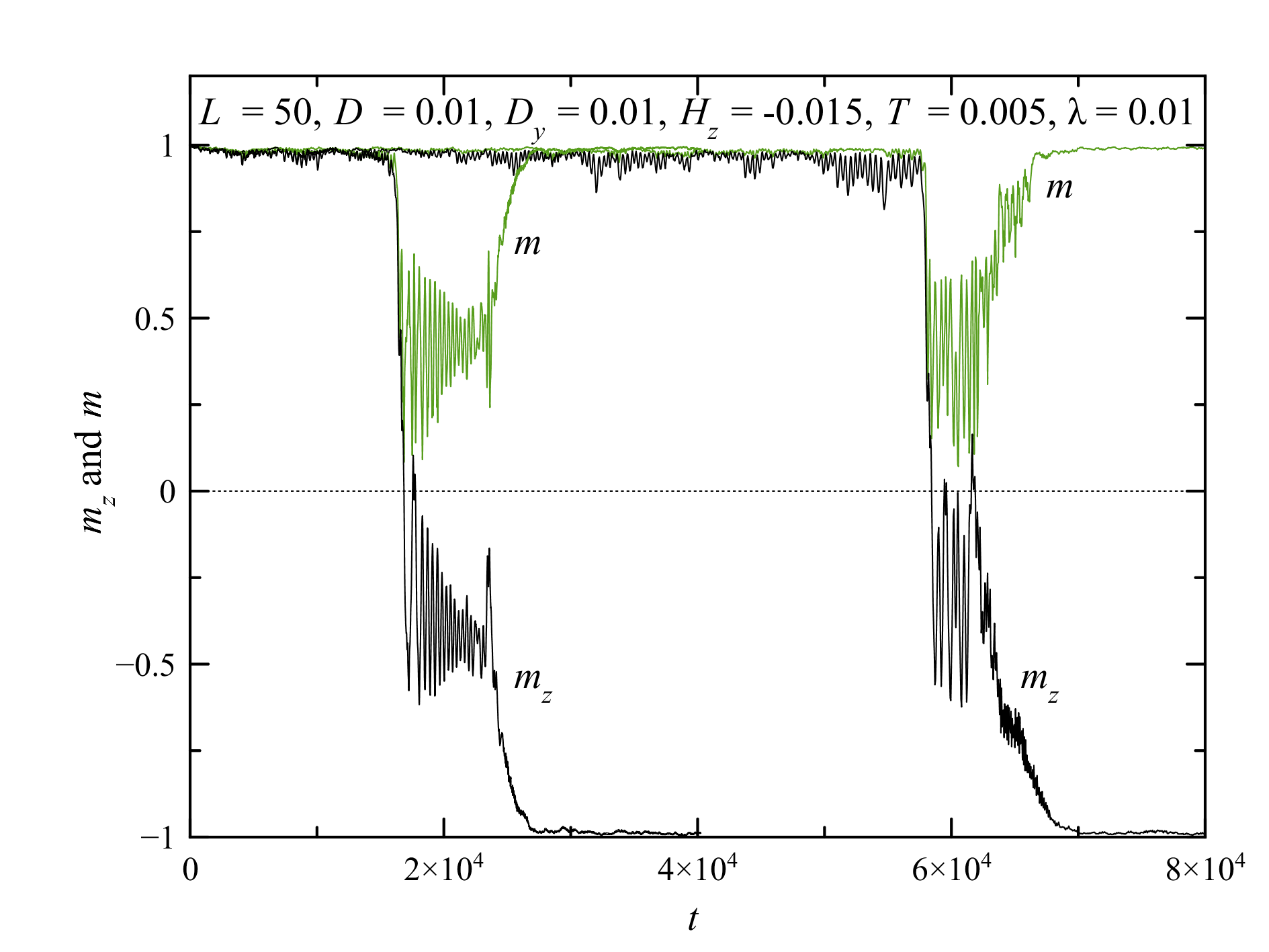}
\par\end{centering}
\begin{centering}
\includegraphics[width=9cm]{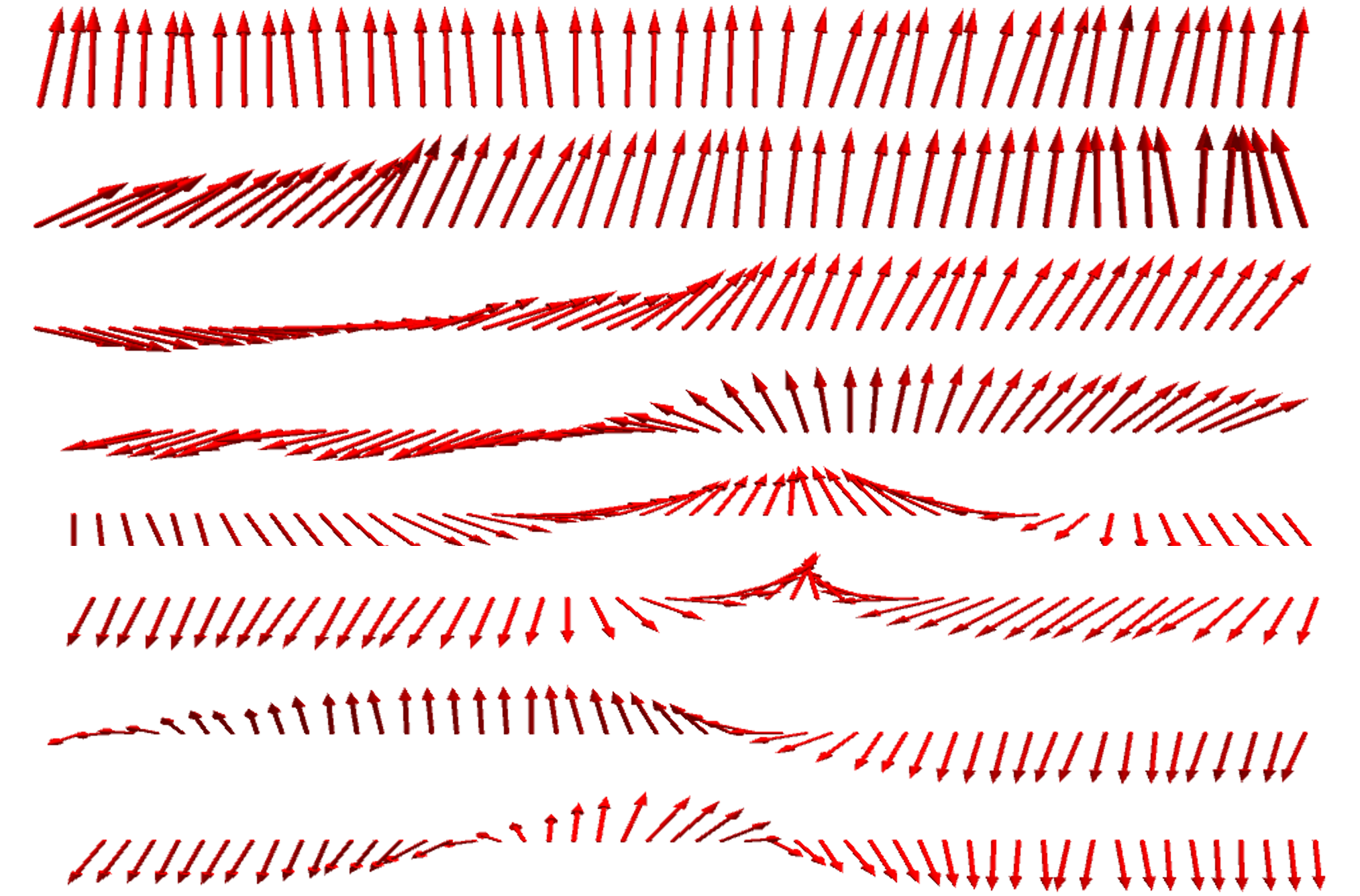}
\par\end{centering}
\caption{Two realizations of thermal switching of a field-biased biaxial magnetic
chain of $50$ spins with fast and unstable domain-wall motion. Upper
panel: time dependence of the magnetization. Lower panel: spin configurations
at different moments of switching for the second realization. (Easy
axis vertical). }

\label{Fig_Spins_switching_stages_L=00003D50_Dz=00003DDy=00003D0.01_Hz=00003D-0.015_lam=00003D0.01}
\end{figure}

Field-biased chains show more affinity to crossing the barrier via
moving domain walls. Thermal switching dynamics of a 100-spins chain
with $D=0.01$ and $H_{z}=-0.015$ in Fig. \ref{Fig_mz_vs_t_L=00003D100_Dz=00003D0.01_Dy=00003D0,0.01_Hz=00003D-0.015}
shows a significant reduction of $m$ on barrier crossing. After emerging
at one of the chain's ends, the DW is pushed to the other end by the
bias field $H_{z}.$ In the axially-symmetric case, domain walls can
move only via the damping with the mobility $v/H_{z}\propto\lambda$,
so that for $\lambda=0.01$ their speed $v$ is low, as can be seen
on the right side of Fig. \ref{Fig_mz_vs_t_L=00003D100_Dz=00003D0.01_Dy=00003D0,0.01_Hz=00003D-0.015}.
During this slow motion, the spins in the DW are precessing around
$z$ axis.

Adding the hard-axis $y$ anisotropy $D_{y}$ breaks the axial invariance
and allows the domain wall to travel with the mobility $v/H_{z}\propto1/\lambda$,
if the applied field is not too strong. In this regime, its speed
is limited by the Walker velocity $v_{W}\propto D_{y}$. In Fig. \ref{Fig_mz_vs_t_L=00003D100_Dz=00003D0.01_Dy=00003D0,0.01_Hz=00003D-0.015}
one can see a fast switching in the chain with $D_{y}=0.01$. Decreasing
$\lambda$ to 0.001 does not change the slope of $m_{z}$ for the
biaxial spin chain, thus one can conclude that the motion of domain
walls here is ballistic. This ballistic motion can become unstable
as shown in Fig. \ref{Fig_Spins_switching_stages_L=00003D50_Dz=00003DDy=00003D0.01_Hz=00003D-0.015_lam=00003D0.01}
that leads to precession of spins around $z$ axis (vertical axis
in this figure) and slowing down the DW motion. This typically causes
formation of solitons (360$^{\circ}$ domain walls) captured inside
the chain, seen in the lower panel of Fig. \ref{Fig_Spins_switching_stages_L=00003D50_Dz=00003DDy=00003D0.01_Hz=00003D-0.015_lam=00003D0.01},
that take some time to exit through the ends. The dynamics of a chain
of a finite length $L_{z}$ becomes more complicated because of the
boundary conditions at the ends which influence the domain walls and
cause reflection of spin waves. The initial moment of reversal seen
in the second, third, and fourth rows looks like a high-amplitude
spin wave turning spins on the left perpendicular to the easy axis
rather than like a nascent domain wall.

\begin{figure}
\begin{centering}
\includegraphics[width=9cm]{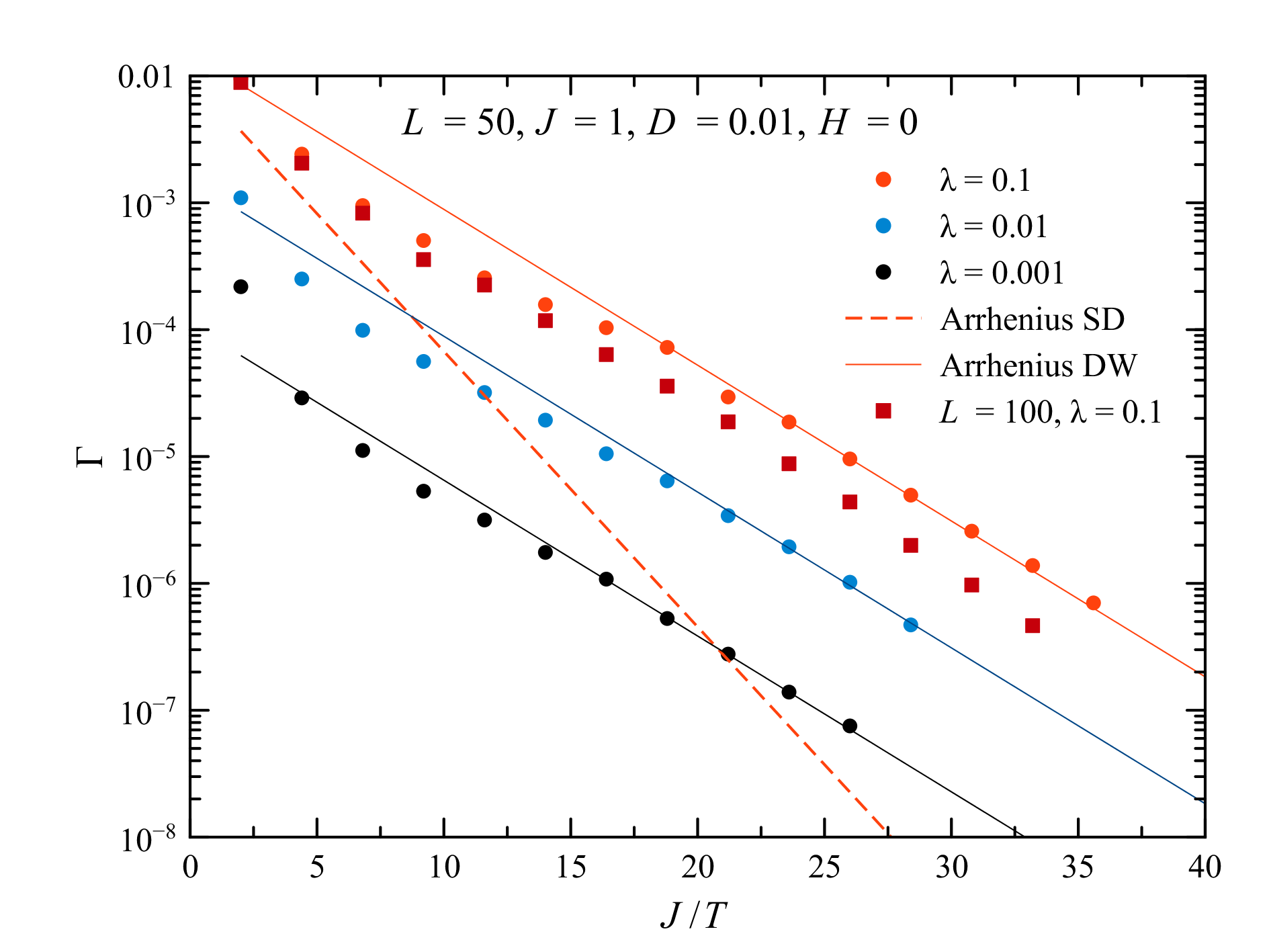}
\par\end{centering}
\begin{centering}
\includegraphics[width=9cm]{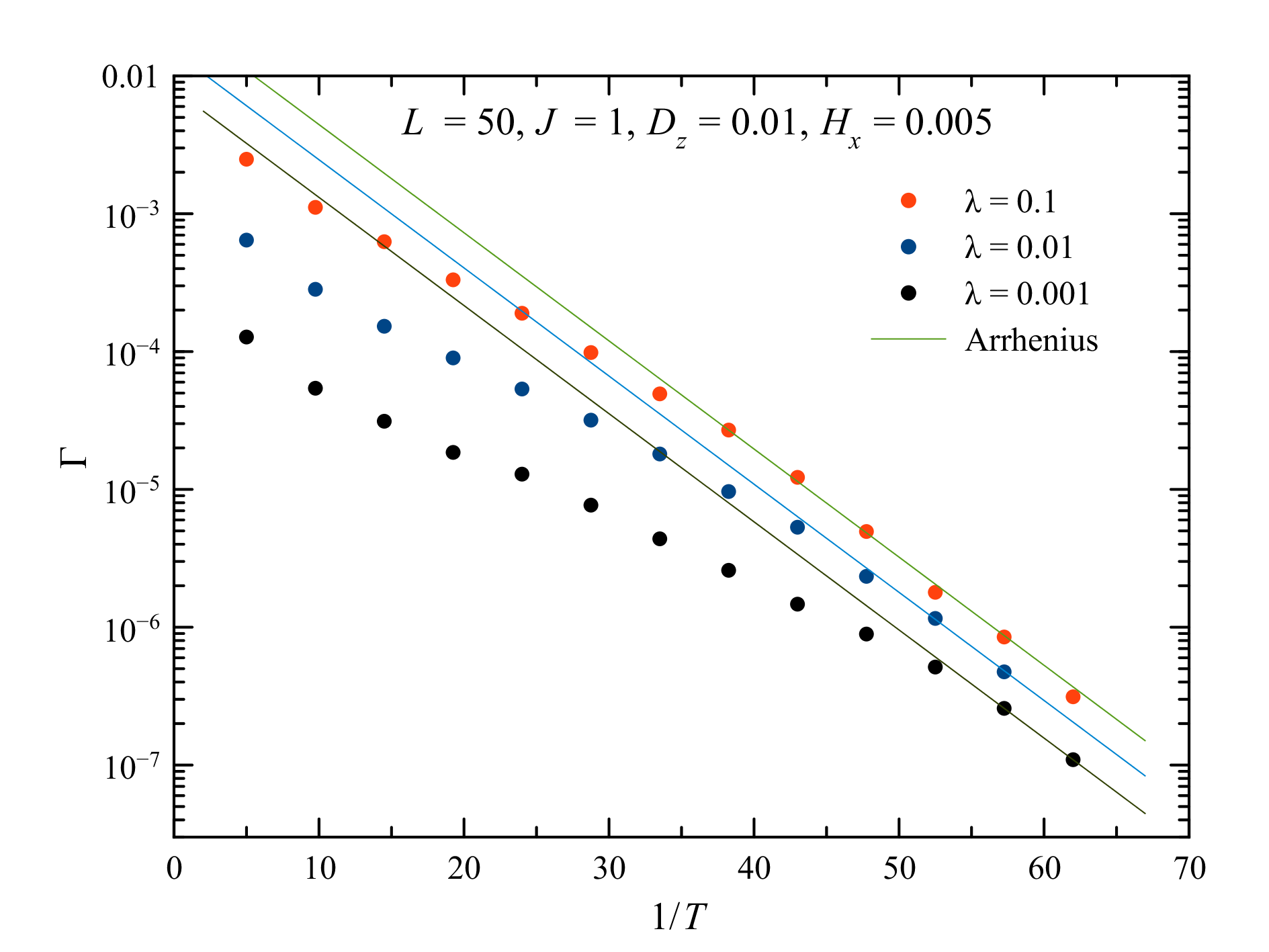}
\par\end{centering}
\caption{Thermal escape rate for the 50-spin chain with $D=0.01$ and $\lambda=0.1$,
0.01, and 0.001. Upper panel: zero field. Results for the 100-spin
chain with $\lambda=0.1$ added for a comparison. Lower panel: $H_{x}=0.05$.
For $\lambda=0.001$, the asymptotic behavior is realized only at
very low temperatures.}

\label{Fig_Gamma_vs_Tinv_L=00003D50_Dz=00003D0.01_H=00003D0_lam=00003D0.1,0.01,0.001}
\end{figure}

Computations of the temperature dependence of the thermal activation
rates were performed on the 50-spin chain with $D=0.01$ and the damping
values $\lambda=0.1$, 0.01, and 0.001. In the axially-symmetric case
shown in the upper panel of Fig. \ref{Fig_Gamma_vs_Tinv_L=00003D50_Dz=00003D0.01_H=00003D0_lam=00003D0.1,0.01,0.001}
the Arrhenius regime with the non-uniform barrier given by Eq. (\ref{U_DW})
sets in at low temperatures. The prefactor $\Gamma_{0}$ is clearly
proportional to $\lambda$, as it is for single-domain particles in
this case. Data for the 100-spin chain show a slower escape with an
apparent barrier slightly higher at low temperatures. This can indicate
an increasing contribution of the nucleation via a couple of opposite
domain walls inside the chain for which the barrier is twice as large

The data in the case of a strong transverse field in the lower panel
of Fig. \ref{Fig_Gamma_vs_Tinv_L=00003D50_Dz=00003D0.01_H=00003D0_lam=00003D0.1,0.01,0.001}
align with the Arrhenius dependence with the barrier given by Eq.
(\ref{U_DW_hx}), the energy of a domain wall with transverse field.
However, this happens at rather low temperatures, especially in the
low-damping case. This should not be a surprise as the dynamics of
domain walls crossing the barrier is rather complicated, as we have
seen above. 

\section{Linear instability of the uniform barrier state\label{sec:Linear-instability} }

Let us investigate the stability of the single-domain barrier state
with spins directed along $\mathbf{e}_{b}$ \textendash{} the barrier
direction in the $xz$ plane. Taking into account deviations from
the barrier direction, the spins can be representes as
\begin{equation}
\mathbf{s}=\mathbf{e}_{b}\sqrt{1-\psi^{2}}+\mathbf{e}_{b'}\psi\cong\mathbf{e}_{b}\left(1-\frac{1}{2}\psi^{2}\right)+\mathbf{e}_{b'}\psi,\label{spin_deviation-1-1}
\end{equation}
where $\mathbf{e}_{b'}$ is in the $xz$ plane perpendicular to $\mathbf{e}_{b}$.
Substituting this into Eq. (\ref{E_prepared}), one obtains the nonuniform
energy up to $\psi^{2}$: 
\begin{equation}
\delta E=\frac{\mathcal{N}D}{L_{z}}\intop_{0}^{L_{z}}dz\left\{ \delta^{2}\psi'^{2}-2A\psi-B\psi^{2}\right\} ,\label{delta_E_psi}
\end{equation}
where
\begin{eqnarray}
A & \equiv & \left(\mathbf{e}_{z}\cdot\mathbf{e}_{b}\right)\left(\mathbf{e}_{z}\cdot\mathbf{e}_{b'}\right)+\mathbf{h}\cdot\mathbf{e}_{b'}\nonumber \\
B & \equiv & \left(\mathbf{e}_{z}\cdot\mathbf{e}_{b'}\right)^{2}-\left(\mathbf{e}_{z}\cdot\mathbf{e}_{b}\right)^{2}-\mathbf{h}\cdot\mathbf{e}_{b}>0,
\end{eqnarray}
Here the linear term must vanish if the barrier direction is chosen
properly: $A=0$. This defines $\theta_{b}$ in $\left(\mathbf{e}_{z}\cdot\mathbf{e}_{b}\right)=\cos\theta_{b}$
and $\left(\mathbf{e}_{z}\cdot\mathbf{e}_{b'}\right)=\sin\theta_{b}$.
For an arbitrarily directed $\mathbf{h}$, there is no analytical
solution for $\theta_{b}$, although it can be obtained numerically.
Analytically solvable cases are
\begin{equation}
h_{z}=0,\qquad\theta_{b}=\pi/2,\qquad B=1-h_{x}
\end{equation}
and
\begin{equation}
h_{x}=0,\qquad\cos\theta_{b}=-h_{z},\qquad B=1-h_{z}^{2}.
\end{equation}

One has to consider small non-uniform deviations from the barrier
state, satisfying the boundary conditions $s_{\alpha}'=0$ and orthogonal
to a constant, and check whether they can reduce the energy. In fact,
one can use the Fourier series for $\psi(z)$. The most dangerous
perturbation is $\psi(z)=p\cos\left(\pi z/L_{z}\right)$. Substituting
it into Eq. (\ref{delta_E_psi}), one obtains 
\begin{equation}
\delta E=\frac{1}{2}\mathcal{N}Dp^{2}\left(\frac{\pi^{2}\delta^{2}}{L_{z}^{2}}-B\right).
\end{equation}
This is positive and thus the SD state is stable for $L_{z}<\pi\delta/B$
that yields
\begin{equation}
L_{z}<\pi\delta\begin{cases}
/\sqrt{1-h_{x}}, & h_{z}=0\\
/\sqrt{1-h_{z}^{2}}, & h_{x}=0.
\end{cases}
\end{equation}

For the periodic boundary conditions (e.g., for magnetic rings), the
dangerous perturbation is $\psi=p\cos\left(2\pi z/L_{z}\right)$ that
increases the exchange energy by the factor of four and makes the
nonuniform solution more expensive. In this case the SD state is stable
for $L_{z}<2\pi\delta/B$. 

For a particle of a cubic shape, there are similar instabilities with
nonuniformities along the $x$, $y$, and $z$ spatial axes. Nonuniformities
along two or three axes at the same time cost too much exchange energy
and are not viable. However, for larger particles, another mechanism,
the nucleation near the surface, becomes important \cite{hinnow98prb}.

\begin{figure}
\begin{centering}
\includegraphics[width=9cm]{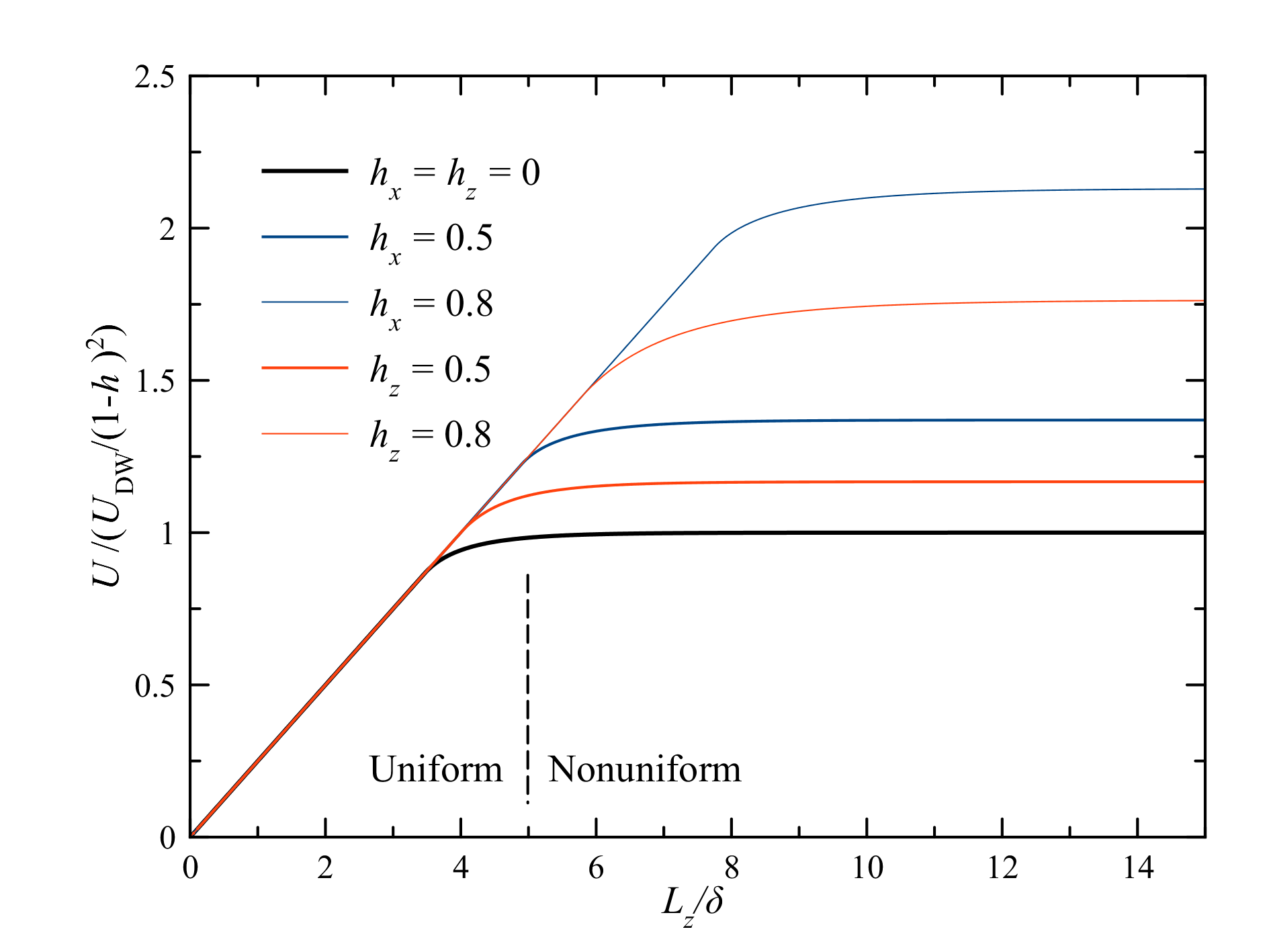}
\par\end{centering}
\caption{Uniform-nonuniform crossover of the energy barrier on the magnetic
particle's length $L_{z}$, relative to the domain-wall width $\delta$.}

\label{Fig_U_vs_K}
\end{figure}

\section{Variational solution in the transition region\label{sec:Variational-solution}}

To qualitatively study the transition between the uniform and nonuniform
thermal activation, one can use the variational approach. In the simplest
case of a zero field, the variational $Ansatz$ has the form

\begin{equation}
s_{z}=\tanh\left[p\left(z-z_{0}\right)\right],\qquad s_{x}=1/\cosh\left[p\left(z-z_{0}\right)\right],\label{DW_profile_hx=00003D0_variational}
\end{equation}
where $p$ is a variable parameter. These functions do not satisfy
the boundary conditions. However, since the boundary conditions follow
from the energy minimization, within the variational approach they
can be discarded. The energy in our model, Eq. (\ref{E_prepared}),
for $\mathbf{h}=0$ becomes
\begin{equation}
E=\mathcal{N}D\frac{1}{L_{z}}\intop_{0}^{L_{z}}dz\frac{\delta^{2}p^{2}+1}{\cosh^{2}\left[p\left(z-z_{0}\right)\right]}.
\end{equation}
Integration yields
\begin{equation}
E=\mathcal{N}D\frac{1+\delta^{2}p^{2}}{pL_{z}}\left[\tanh\left[p\left(L_{z}-z_{0}\right)\right]+\tanh\left[pz_{0}\right]\right].\label{Ham_variational_z0}
\end{equation}
that has a flat maximum at $z_{0}=L_{z}/2$, DW in the center of the
particle. This corresponds to the barrier height
\begin{equation}
U(p)=2\mathcal{N}D\frac{1+\delta^{2}p^{2}}{pL_{z}}\tanh\frac{pL_{z}}{2}.\label{U(p)}
\end{equation}
This expression has to be minimized with respect to $p$. For $p\rightarrow0$
this result recovers the barrier for the single-domain particle, $U_{SD}=\mathcal{N}D$.
For $L_{z}\gg\delta$ the solution satisfies $pL_{z}\gg1$, so that
$\tanh\left(pL_{z}/2\right)\cong1$. Then minimization yields $p=1/\delta$
and $U$ given by Eq. (\ref{U_DW}).

To study the transition between the two regimes, one can expand $U(p)$
as
\begin{equation}
\frac{U(q)}{\mathcal{N}D}\cong1+\left(1-\frac{L_{z}^{2}}{12\delta^{2}}\right)\left(p\delta\right)^{2}+\frac{L_{z}^{2}}{12\delta^{2}}\left(-1+\frac{L_{z}^{2}}{10\delta^{2}}\right)\left(p\delta\right)^{4}.\label{U(q)_small_q}
\end{equation}
Thus, the instability of the single-domain state within the variational
approach sets in at $L_{z}=L_{z,c}=\sqrt{12}\delta\simeq3.46\delta$
that differs from the exact criterion above by 10\%. For $L_{z}>L_{z,c}$,
the coefficient in front of the fourth-order term is positive, thus
this is a second-order transition within the variational approach.

Using the similar method, one can study the transition from uniform
to nonuniform barrier in the presence of transverse and longitudinal
fields. Fig. \ref{Fig_U_vs_K} shows the barrier vs $L_{z}$, relative
to the domain-wall width $\delta$. In the uniform region, the barrier
grows linearly with the particle's volume, then it crosses over to
a plateau. Transverse and longitudinal fields favor the uniform state,
thus in the field, transition between the regimes happens at larger
$L_{z}$. 

\section{Discussion\label{sec:Discussion}}

It was shown that in the realistic case of weak damping, the pulse-noise
approach to the solution of the Landau-Lifshitz-Langevin equation
for spin systems coupled to a heat bath is efficient and can be used
to compute thermally-activated escape rates of magnetic nanoparticles
in uniform and nonuniform switching regimes. It was found that single-domain
magnetic particles show much higher switching rates than the equivalent
single-spin models because of the internal thermal disordering manifesting
itself in the temperature dependence of the effective anisotropy constant.
The temperature dependence of the energy barrier was not a secret
and it was understood that the apparent barriers extracted from the
Arrhenius switching rates are the barriers at $T=0$. The temperature
corrections to the barriers strongly change the prefactors, so that
a comparison with theoretical expressions is problematic. 

Since thermal disordering of magnetic particles proves to be very
important in their thermal switching, it is worth trying to take into
account quantum-mechanical effects in the temperature dependence of
the magnetization (e.g., the Bloch's law). While obtaining a quantum-mechanical
expression for the particle's magnetization from the quantum spin-wave
theory seems to be possible, it is unclear how to make first-principle
computations with quantum mechanics and Langevin fields.

It was demonstrated that for strong anisotropy at high temperature
the magnetization switching is predominantly longitudinal with the
magnetization reducing to zero at the barrier crossing. This is neither
a uniform rotation nor a nonuniform rotation since at high temperatures
the neighboring spins strongly deviate from collinearity in a random
way.

At lower temperatures, the crossover from uniform (single-domain)
rotation to nonuniform rotation by surmounting the barrier has been
studied analytically, including the linear instability boundary of
the uniform barrier state. In the region of nonuniform rotation analytical
expressions for the barriers in the presence of transverse and longitudinal
field have been worked out. Computations using the pulse-noise method
are in a good accordance with the values of the barriers, although
in some cases the Arrhenius dependence with the given barrier sets
in at pretty low temperatures (such as the curve $\lambda=0.001$
in the lower panel of Fig. \ref{Fig_Gamma_vs_Tinv_L=00003D50_Dz=00003D0.01_H=00003D0_lam=00003D0.1,0.01,0.001}).
\begin{acknowledgments}
This work has been supported by Grant No. DE-FG02-93ER45487 funded
by the US Department of Energy, Office of Science.
\end{acknowledgments}

\section*{Appendix: Extracting escape rates from escape data}

\begin{figure}
\begin{centering}
\includegraphics[width=9cm]{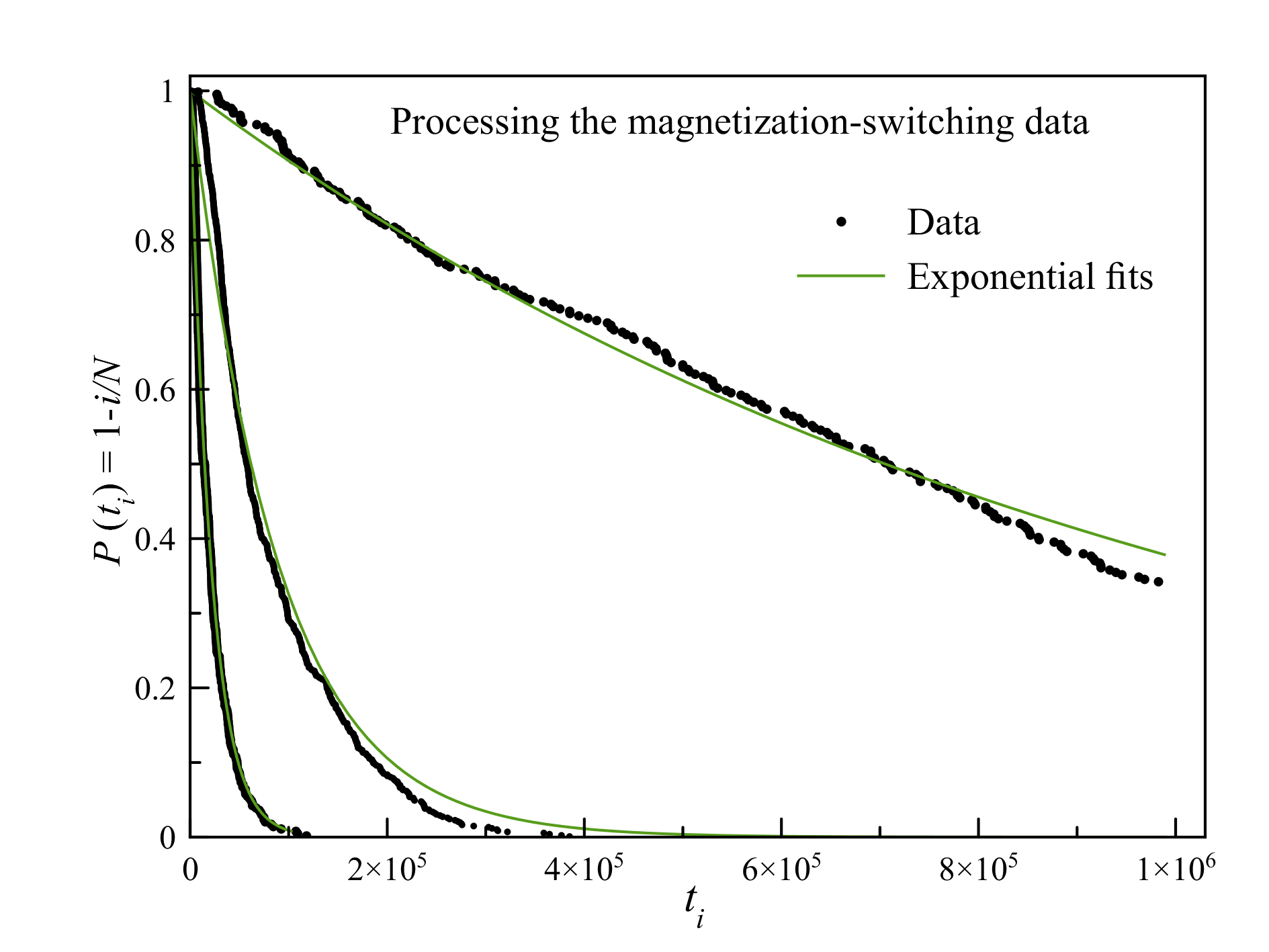}
\par\end{centering}
\caption{Extracting escape rates from escape data. This method does not require
all particles to escape.}

\label{Fig_PStay_vs_t}
\end{figure}

The usual way to numerically compute escape rates is to run the numerical
evolution routine until the particle crosses the barrier or using
another stopping criterion like $m_{z}=0$ and record the (first)
passage time $t_{i}$ for each $i$th run. The inverse of the mean
first-passage time (MFPT), averaged over many runs, is identifies
with the escape rate, $\Gamma=1/\left\langle t_{i}\right\rangle $. 

The mathematics behind this relies on the assumption that the probability
for a particle to stay in the initial state decreases exponentially
with time: $P(t)=e^{-\Gamma t}$. This is the unnormalized distribution
of passage times, so that the mean passage time is given by $\mathrm{MFPT}=\Gamma\intop_{0}^{\infty}dt\,tP(t)=1/\Gamma$,
where $\Gamma$ in front of the integral normalizes the distribution.
This confirms the validity of the standard method of extracting $\Gamma$.

A drawback of this method to extract $\Gamma$ is that absolutely
all particles must escape. If the preset computation time was too
short and not all particles excaped, the efforts have been wasted
and the computation has to be repeated. Increasing the preset computation
time leads to very long computations at the lowest temperatures as
the distribution of passage times becomes broad and there are extremely
long-lived particles.

A more advanced method is to consider the sorted list of passage times
$t_{i}$ and identify $i$ with the number of particles already escaped
by the time $t_{i}$. Then the probability to escape by the time $t_{i}$
becomes $i/N$, where $N$ is the total number of runs used in the
computation, i.e., the total number of particles. Thus the staying
probability at $t_{i}$ can be expressed as 
\begin{equation}
P(t_{i})=1-\frac{i}{N}.
\end{equation}
Assuming that $P(t)$ is exponential, one can fit the list $\left\{ t_{i},P(t_{i})\right\} $
with the exponential function to extract $\Gamma$. Since fitting
is a nonlinear procedure, it can fail, thus it is better to be avoided.
Instead of the fitting, one can extract $\Gamma$ by resolving the
exponential as $\Gamma=-\ln P(t)/t$. This results in the formula
used in this work, as well as in Ref. \cite{gar17pre}:
\begin{equation}
\Gamma=-\left\langle \frac{1}{t_{i}}\ln\left(1-\frac{i}{N}\right)\right\rangle .
\end{equation}
This formula gives practically the same values of $\Gamma$ as the
exponential fits shown in Fig. \ref{Fig_PStay_vs_t}. Here the averaging
includes only those particles that escaped, whereas $N$ is the total
number of particles. This formula is robust and does not require all
particles to escape. To the contrary, limiting the preset computation
time allows to speed up the computations at the lowest temperatures
extracting the information from the subset of the shortest-lived particles.
The latter amounts to determining the exponential by its initial part,
as it can be seen in Fig. \ref{Fig_PStay_vs_t}. 

In these computations, the preset computation time was $t_{\max}=10^{6}$.
With the standard method of data processing, this would allow to compute
escape rates significantly higher than $1/t_{\max}$, say, down to
$\Gamma=10^{-5}$. With the current method, escape rates down to $\Gamma=10^{-7}$
have been computed.

Another advantage of this method is the possibility of plotting the
staying probability, as it is done in Fig. \ref{Fig_PStay_vs_t} to
check whether it is exponential. In particular, at high temperatures
this curve is more resembling a Gaussian, thus the extracted $\Gamma$
are only approximate.

\bibliographystyle{apsrev4-1}
\bibliography{C:/BIBLIOTHEK/gar-own,C:/BIBLIOTHEK/gar-books,C:/BIBLIOTHEK/gar-relaxation,C:/BIBLIOTHEK/gar-spin,C:/BIBLIOTHEK/gar-superparamagnetic,C:/BIBLIOTHEK/gar-oldworks,C:/BIBLIOTHEK/gar-general}

\end{document}